# Stationary solutions of the second-order equation for fermions in Kerr-Newman space-time

V.P.Neznamov*, I. I. Safronov, V.Ye. Shemarulin

[1]FSUE "RFNC-VNIIEF", Russia, Sarov, Mira pr., 37, 607188

## Abstract

When using the quantum mechanical second-order equation with the effective potential of the Kerr-Newman (KN) field for fermions, results were obtained that qualitatively differ from results obtained when using the Dirac equation.

In presence of two event horizons, existence of degenerate stationary bound states was proved for charged and uncharged fermions with square integrable wave functions vanishing on event horizons. The fermions in such states are localized near the event horizons with the maxima of probability densities away from the event horizons by fractions of the Compton wave length of fermions versus the values of coupling constants, the values of angular and orbital momenta $j, l$ and the value of the azimuthal quantum number $m_\varphi$.

In the case of extreme KN fields, absence of stationary bound states of fermions was shown for any values of coupling constants.

Existence of discrete energy spectra was shown for charged and uncharged fermions in the field of naked KN singularity at definite values of physical parameters.

The KN naked singularity poses no threat to cosmic censorship because of the regular behavior of the effective potentials of the KN field in quantum mechanics with the second-order equation.

*E-mail: vpneznamov@vniief.ru; vpneznamov@mail.ru

## 1. Introduction

In 1976, Chandrasekhar [1], [2], Page [3] and Toop [4] showed that it is possible to separate angular and radial variables in the Dirac equation for the point fermion in the Kerr and Kerr-Newman space-time. Since then, the behavior of the fermion in the external Kerr and Kerr-Newman fields has been studied in many papers (see, for example, [7] - [18]). In [19] - [22], the Dirac equation was studied in the Kerr-Newman space-time at $G \to 0$, where $G$ is the Newtonian universal gravitational constant.

The nontrivial topology of Kerr and Kerr-Newman metrics associated with ring singularity was studied in [23], [24] and later in [21], [22]. It is shown in [23], [24] that the maximally analytically extended Kerr-Newman manifold cross-linked through the ring. Such a topology persists for the Kerr metric and for "zero" gravitation $G \to 0$.

In presence of event horizons, absence of stationary bound states of the Dirac equation is shown in [8] - [10], [15] for the domains of wave functions of the fermion outside and under event horizons.These conclusions are also validated in this paper in section 2.

The situation qualitatively changes if the motion of fermions is described by a self-conjugate second-order equation with the spinor wave function. The second-order equation in the external electromagnetic field was first proposed by Dirac [25]. Using the ratio between the upper and lower spinors of the Dirac bispinor, the second-order equation can be written as two separate equations with spinor wave functions. In this case, for self-conjugacy of second-order equations, it is necessary to perform appropriate nonunitary similarity transformations for each of them (see, for example, [26]). As a result, new physical consequences can arise when using self-conjugate second-order equations with spinor wave functions in quantum mechanics of half-spin particles in external electromagnetic and gravitational fields.

In this paper, the second-order equation with the effective potential is used to describe the motion of fermions in Kerr-Newman (KN) geometry. This equation also describes the motion of fermions in Schwarzschild [27] and Reissner-Nordström fields [28] at appropriate values of initial parameters. In absence of gravitation, the second-order equation describes the motion of fermions in the effective Coulomb field [26].

In presence of two event horizons of the KN metric for charged and uncharged fermions, we proved the existence of stationary bound states of fermions with square-integrable wave functions vanishing on event horizons. The fermions in such states are localized near event horizons. The maxima of probability densities of particle detection are away from the event horizon by several fractions of the Compton wavelength of fermions versus the values of gravitational and electromagnetic coupling constants, the value of angular momentum of the KN field source, the value of quantum numbers of angular and orbital momenta $j, l$ and the azimuthal quantum number $m_\varphi$.

The impossibility of existence of stationary bound states of fermions was proved for the extreme KN field with the single event horizon.

Existence of discrete energy spectra of half-spin particles was validated and shown numerically for the naked KN singularity at definite values of physical parameters.

In Kerr-Newman geometry, the effective potential of the second-order equation is finite at the origin. Hence, the KN field of naked singularity in quantum mechanics of fermion motion does not pose any threat to cosmic censorship. Similar conclusions were made earlier for the Reissner-Nordström naked singularity [28] and for a number of timelike naked singularities in quantum mechanics of spinless particles [29].

The paper is arranged as follows. In section 2, the self-conjugate Dirac equation is obtained for KN geometry, the variables are separated, the system of equations is presented for radial wave functions, the asymptotics of radial wave functions are considered, nonregular stationary solutions are determined for the Dirac equation in the KN field. The examination is carried out in presence of two event horizon, for the extreme KN field with the single event horizon and for the case of KN naked singularity.

In section 3, in the Kerr-Neman geometry, the self-conjugate second-order equation with spinor wave functions of fermions is obtained. The singularities of the effective potential and asymptotics of the radial wave function of the second-order equation are discussed.

The singularities of effective potentials and the behavior of radial wave functions testify to possibility of existence of stationary bound states of fermions both in presence of two horizons and for the case of KN naked singularity. Absence of the potential well for the extreme KN field testifies to absence of bound states of fermions in this case.

In section 4, the second-order equation with the effective potential after the Prüfer transformation [30] - [33] and introduction of a phase function is presented as a system of non-linear differential first-order equations.

Section 5 presents the results of numerical calculations for determination of stationary bound states of fermions in the KN field. The procedure of calculations in the neighborhood of singular points ( $r = \pm\infty$ are regular singular points, $r = (r_\pm)_{KN}$, $r = r_{cl}$ are nonregular singular points) is discussed. Here, $(r_\pm)_{KN}$ are radii of outer and inner event horizons, $r_{cl}$ is the radius of the impenetrable barrier, which can emerge at definite values of physical parameters near the origin.

In section 6, absence of a threat to cosmic censorship in quantum mechanics of fermion motion in the KN field of naked singularity is established.

In the Conclusions, the basic results of the paper are presented and discussed.

In the Appendix, the explicit form of effective potentials of the second-order equation in the KN field is presented.

## 2. Dirac equations in the Kerr and Kerr-Newman fields

In the paper, we generally use the system of units $\hbar = c = 1$; the signature of the metric of the Minkowsky space is chosen to be

$$\eta_{\underline{\alpha\beta}} = \mathrm{diag}\left[1,-1,-1,-1\right]. \tag{1}$$

In (1) and below, the underlined indexes are local. The indexes with Greek letters assume the values of $0,1,2,3$; the indexes with Latin letters assume the values of $1,2,3$. The standard rule for summation over repeated indexes is used.

### 2.1. Kerr and Kerr-Newman metrics

The stationary Kerr-Newman metric is characterized by a point source of mass $M$ and charge $Q$, rotating with the angular momentum of $\mathbf{J} = Mc\mathbf{a}$. The Kerr-Newman metric in Boyer-Lindquist coordinates $\left(t, r, \theta, \varphi\right)$ [34] can be presented as

$$\begin{aligned} ds^2 = {} & \left(1 - \frac{r_0 r - r_Q^2}{r_K^2}\right) dt^2 + \frac{2a\left(r_0 r - r_Q^2\right)}{r_K^2} \sin^2\theta dt d\varphi - \frac{r_K^2}{\Delta_{KN}} dr^2 - r_K^2 d\theta^2 - \\ & - \left(r^2 + a^2 + \frac{a^2\left(r_0 r - r_Q^2\right)}{r_K^2} \sin^2\theta\right) \sin^2\theta d\varphi^2. \end{aligned} \tag{2}$$

In (2), $r_K^2 = r^2 + a^2\cos^2\theta$; $\Delta_{KN} = r^2 f_{KN} = r^2\left(1 - \dfrac{r_0}{r} + \dfrac{r_Q^2 + a^2}{r^2}\right)$; $r_0 = 2GM/c^2$ is the gravitational radius of the Schwarzschild field; $r_Q = \sqrt{G}Q/c^2$; $G$ is the gravitational constant, $c$ is the speed of light.

1. If $r_0 > 2\sqrt{a^2 + r_Q^2}$, then

$$f_{KN} = \left(1 - \frac{\left(r_+\right)_{KN}}{r}\right)\left(1 - \frac{\left(r_-\right)_{KN}}{r}\right), \tag{3}$$

where $\left(r_\pm\right)_{KN}$ are the radii of the inner and outer event horizons of the Kerr-Newman field,

$$\left(r_\pm\right)_{KN} = \frac{r_0}{2} \pm \sqrt{\frac{r_0^2}{4} - a^2 - r_Q^2}. \tag{4}$$

2. The case of $r_0 = 2\sqrt{a^2 + r_Q^2}$, $(r_+)_{KN} = (r_-)_{KN} = r_0/2$ corresponds to the extreme Kerr-Newman field.

3. The case of $r_0 < 2\sqrt{a^2 + r_Q^2}$ corresponds to naked singularity of the Kerr-Newman field. In this case, $f_{KN} > 0$.

4. At $Q = 0$, the Kerr-Newman metric transforms to the Kerr metric with

$$\Delta_K = r^2 f_K = r^2 \left(1 - \frac{r_0}{r} + \frac{a^2}{r^2}\right), \tag{5}$$

$$(r_\pm)_K = \frac{r_0}{2} \pm \sqrt{\frac{r_0^2}{4} - a^2}. \tag{6}$$

**2.2. Self-conjugate Dirac equation**

If we write the Dirac equation in Hamiltonian form, then Chandrasekhar [1], [2] and Page [3] Hamiltonians will be pseudo-Hermitian with Parker weight factors [35] in scalar products of wave functions. We can obtain self-conjugate Hamiltonians with plane scalar products of wave functions by using methods of pseudo-Hermitian quantum mechanics [36] - [38]. Such Hamiltonians were obtained in [38], [39] for the Kerr and Kerr-Newman metrics. In [40], equivalence of the self-conjugate Hamiltonian in [38] and the Chandrasekhar Hamiltonian in [1], [2] was proved.The drawback of Hamiltonian in [38] is impossibility to separate variables. Below, self-conjugate Dirac equations, allowing separation of variables, are obtained for the Kerr and Kerr-Newman metrics.

Let us begin with Dirac equations for Kerr and Kerr-Newman metrics obtained in the papers by Chandrasekhar and Page and written in the form of [9]

$$(G-1)\Psi = \left(\begin{array}{c|c|c|c} 1 & 0 & \alpha_+ & \beta_+ \\ \hline 0 & -1 & \beta_- & \hat{\varepsilon}(\Delta_{KN})\alpha_- \\ \hline \hat{\varepsilon}(\Delta_{KN})\bar{\alpha}_- & -\bar{\beta}_+ & -1 & 0 \\ \hline -\bar{\beta}_- & \bar{a}_+ & 0 & 1 \end{array}\right)\Psi = 0. \tag{7}$$

In (7), $\Psi$ is the bispinor wave function.

$$\beta_\pm = \frac{1}{\sqrt{\rho_K^2}}\left(i\frac{\partial}{\partial\theta} + i\frac{\operatorname{ctg}\theta}{2} + \frac{\alpha_a \sin\theta}{2\rho_K^2}(\rho - i\alpha_a\cos\theta)\right) \pm \frac{1}{\sqrt{\rho_K^2}}\left(\alpha_a \sin\theta\frac{\partial}{\partial t} + \frac{1}{\sin\theta}\frac{\partial}{\partial\varphi}\right), \tag{8}$$

$$\bar{\beta}_\pm = \frac{1}{\sqrt{\rho_K^2}}\left(i\frac{\partial}{\partial\theta} + i\frac{\operatorname{ctg}\theta}{2} - \frac{\alpha_a \sin\theta}{2\rho_K^2}(\rho + i\alpha_a\cos\theta)\right) \pm \frac{1}{\sqrt{\rho_K^2}}\left(\alpha_a \sin\theta\frac{\partial}{\partial t} + \frac{1}{\sin\theta}\frac{\partial}{\partial\varphi}\right), \tag{9}$$

$$\alpha_{\pm} = -\frac{\hat{\varepsilon}(\Delta_{KN})}{\sqrt{\rho_K^2 |\Delta_{KN}|}} \left( i\left(\rho^2 + \alpha_a^2\right)\frac{\partial}{\partial t} + i\alpha_a \frac{\partial}{\partial \varphi} - \alpha_{em}\rho \right) \pm$$
$$\pm \sqrt{\frac{|\Delta_{KN}|}{\rho_K^2}} \left( i\frac{\partial}{\partial \rho} + i\frac{\rho - \alpha}{2\Delta_{KN}} + \frac{i}{2\rho_K^2}\left(\rho - i\alpha_a \cos\theta\right) \right). \tag{10}$$

$$\bar{\alpha}_{\pm} = -\frac{\hat{\varepsilon}(\Delta_{KN})}{\sqrt{\rho_K^2 |\Delta_{KN}|}} \left( i\left(\rho^2 + \alpha_a^2\right)\frac{\partial}{\partial t} + i\alpha_a \frac{\partial}{\partial \varphi} - \alpha_{em}\rho \right) \pm$$
$$\pm \sqrt{\frac{|\Delta_{KN}|}{\rho_K^2}} \left( i\frac{\partial}{\partial \rho} + i\frac{\rho - \alpha}{2\Delta_{KN}} + \frac{i}{2\rho_K^2}\left(\rho + i\alpha_a \cos\theta\right) \right). \tag{11}$$

Expressions (7) - (11) are written in dimensionless variables

$$\rho = \frac{r}{l_c};\ \varepsilon = \frac{E}{m};\ \alpha = \frac{r_0}{2l_c} = \frac{GMm}{\hbar c} = \frac{Mm}{M_P^2}; \alpha_Q = \frac{r_Q}{l_c} = \frac{\sqrt{G}Qm}{\hbar c} = \frac{\sqrt{\alpha_{fs}}}{M_P} m \frac{Q}{e};$$
$$\alpha_a = \frac{a}{l_c};\ \alpha_{em} = \frac{qQ}{\hbar c} = \alpha_{fs}\frac{qQ}{e^2}. \tag{12}$$

Here, $l_c = \frac{\hbar}{mc}$ is the Compton wave length of a half-spin particle;

$M_P = \sqrt{\frac{\hbar c}{G}} = 2.2 \cdot 10^{-5}$ g $\left(1.2 \cdot 10^{19}\,\text{GeV}\right)$ is the Planck mass; $\alpha_{fs} = \frac{e^2}{\hbar c} \approx \frac{1}{137}$ is the electromagnetic constant of the fine structure; $\alpha, \alpha_{em}$ are the gravitational and electromagnetic coupling constants; $\alpha_Q, \alpha_a$ are dimensionless constants characterizing the source of the electromagnetic field and the ratio of angular momentum $J$ to mass $M$ in the Kerr-Newman metric.

The values of $\rho_K^2, \Delta_{KN}$ in dimensionless variables have the form

$$\rho_K^2 = \rho^2 + \alpha_a^2 \cos^2\theta, \tag{13}$$

$$\Delta_{KN} = \rho^2 f_{KN} = \rho^2 \left( 1 - \frac{2\alpha}{\rho} + \frac{\alpha_a^2 + \alpha_Q^2}{\rho^2} \right). \tag{14}$$

In presence of inner and outer event horizons: $\alpha^2 > \alpha_a^2 + \alpha_Q^2$ and

$$f_{KN} = \frac{\left(\rho - (\rho_+)_{KN}\right)\left(\rho - (\rho_-)_{KN}\right)}{\rho^2}, \tag{15}$$

where

$$(\rho_{\pm})_{KN} = \alpha \pm \sqrt{\alpha^2 - \alpha_a^2 - \alpha_Q^2}. \tag{16}$$

For the extreme KN field, $\alpha^2 = \alpha_a^2 + \alpha_Q^2$, $(\rho_+)_{KN} = (\rho_-)_{KN} = \alpha$, and

$$f_{KN} = \frac{(\rho-\alpha)^2}{\rho^2}. \tag{17}$$

At $\alpha^2 < \alpha_a^2 + \alpha_Q^2$ , the case of naked singularity of the KN field is implemented.

In formulas (10), (11),

$$\hat{\varepsilon}(x) = \begin{cases} 1 & x \ge 0, \\ -1 & x < 0. \end{cases} \tag{18}$$

Let us perform some transformations to obtain a more symmetric form of Eq. (7) in compliance with [9]. Let $S(\rho,\theta)$ and $\Gamma(\rho,\theta)$ be diagonal matrices of the following forms:

$$\begin{gathered} S = \left|\Delta_{KN}^{1/4}\right| \operatorname{diag}\left[(\rho - i\alpha_a\cos\theta)^{1/2}, (\rho - i\alpha_a\cos\theta)^{1/2}, (\rho + i\alpha_a\cos\theta)^{1/2}, (\rho + i\alpha_a\cos\theta)^{1/2}\right], \\ \Gamma = -i\operatorname{diag}\left[(\rho + i\alpha_a\cos\theta), -(\rho + i\alpha_a\cos\theta), -(\rho - i\alpha_a\cos\theta), (\rho - i\alpha_a\cos\theta)\right]. \end{gathered} \tag{19}$$

Then, the transformed wave function

$$\hat{\Psi} = S\Psi \tag{20}$$

satisfies the equation

$$\Gamma S(G-1)S^{-1}\hat{\Psi} = (R+A)\hat{\Psi} = 0, \tag{21}$$

where

$$R = \left(\begin{array}{c|c|c|c} i\rho & 0 & \sqrt{|\Delta_{KN}|}D_+ & 0 \\ \hline 0 & -i\rho & 0 & \hat{\varepsilon}(\Delta_{KN})\sqrt{|\Delta_{KN}|}D_- \\ \hline \hat{\varepsilon}(\Delta_{KN})\sqrt{|\Delta_{KN}|}D_- & 0 & -i\rho & 0 \\ \hline 0 & \sqrt{|\Delta_{KN}|}D_+ & 0 & i\rho \end{array}\right), \tag{22}$$

$$A = \left(\begin{array}{c|c|c|c} -\alpha_a\cos\theta & 0 & 0 & L_+ \\ \hline 0 & \alpha_a\cos\theta & -L_- & 0 \\ \hline 0 & L_+ & -\alpha_a\cos\theta & 0 \\ \hline -L_- & 0 & 0 & \alpha_a\cos\theta \end{array}\right). \tag{23}$$

In (22) and (23),

$$D_\pm = \frac{\partial}{\partial\rho} \mp \frac{1}{\Delta_{KN}}\left((\rho^2 + \alpha_a^2)\frac{\partial}{\partial t} + \alpha_a\frac{\partial}{\partial\varphi} + i\alpha_{em}\rho\right), \tag{24}$$

$$L_\pm = \frac{\partial}{\partial\theta} + \frac{1}{2}\operatorname{ctg}\theta \mp i\left(\alpha_a\sin\theta\frac{\partial}{\partial t} + \frac{1}{\sin\theta}\frac{\partial}{\partial\varphi}\right). \tag{25}$$

Operator $R$ depends only on radial variable $\rho$ and operator $A$ depends only on angular variables $(\theta,\varphi)$.

If we present function $\hat{\Psi}$ as

$$\hat{\Psi}(t,\rho,\theta,\varphi) = e^{-i\varepsilon t} e^{i m_\varphi \varphi} \hat{\Phi}(\rho,\theta), \tag{26}$$

where

$$\hat{\Phi}(\rho,\theta) = \begin{pmatrix} R_-(\rho) S_-(\theta) \\ R_+(\rho) S_+(\theta) \\ R_+(\rho) S_-(\theta) \\ R_-(\rho) S_+(\theta) \end{pmatrix}, \tag{27}$$

then we can obtain the systems of Chandrasekhar-Page equations separately for radial functions $R_\mp(\rho)$ and separately for angular spheroidal harmonics $S_\mp(\theta)$.

Dirac equations (7) and (21) were obtained using Dirac matrics with local indices in spinor representation (Weyl representation)

$$\gamma_W^{\underline{0}} = -\begin{pmatrix} 0 & 1 \\ 1 & 0 \end{pmatrix};\ \gamma_W^{\underline{k}} = \begin{pmatrix} 0 & \sigma^k \\ -\sigma^k & 0 \end{pmatrix};\ \alpha_W^{\underline{k}} = \gamma_W^{\underline{0}} \gamma_W^{\underline{k}} = \begin{pmatrix} \sigma^k & 0 \\ 0 & -\sigma^k \end{pmatrix}. \tag{28}$$

Earlier, we used matrices in Dirac-Pauli representation to record Hamiltonians in Schwarzschild and Reissner-Nordström fields [27], [28]:

$$\gamma_{DP}^{\underline{0}} = \begin{pmatrix} 1 & 0 \\ 0 & -1 \end{pmatrix};\ \gamma_{DP}^{\underline{k}} = \begin{pmatrix} 0 & \sigma^k \\ -\sigma^k & 0 \end{pmatrix};\ \alpha_{DP}^{\underline{k}} = \gamma_{DP}^{\underline{0}} \gamma_{DP}^{\underline{k}} = \begin{pmatrix} 0 & \sigma^k \\ \sigma^k & 0 \end{pmatrix}. \tag{29}$$

The matrices in Dirac-Pauli representation are related to the matices in Weyl representation through unitary transformation

$$\gamma_{DP}^{\underline{\mu}} = M \gamma_W^{\underline{\mu}} M^+;\ M^+ = M^{-1}, \tag{30}$$

$$M = \frac{1}{\sqrt{2}} \begin{pmatrix} \mathrm{I} & \text{-}\mathrm{I} \\ \mathrm{I} & \mathrm{I} \end{pmatrix}. \tag{31}$$

In (31), I is a two-dimensional unity matrix.

Let us perform unitary transformation (31) of wave function (26), (27) and Dirac equation (21). The transformed function $\hat{\Phi}_{DP}$ has the form

$$\hat{\Phi}_{DP} = M \hat{\Phi}(\rho,\theta) = \frac{1}{\sqrt{2}} \begin{pmatrix} (R_- - R_+)\ S_- \\ (R_+ - R_-)\ S_+ \\ (R_- + R_+)\ S_- \\ (R_- + R_+)\ S_+ \end{pmatrix}. \tag{32}$$

Next, let us take into account the property of Chandrasekhar-Page wave functions [21], [41]

$$\begin{aligned} R_-(\rho) &= R_+^*(\rho), \\ R_+(\rho) &= R_-^*(\rho), \end{aligned} \tag{33}$$

and introduce real radial fucntions

$$g(\rho) = R_-(\rho) + R_+(\rho),$$
$$f(\rho) = i\left(R_+(\rho) - R_-(\rho)\right). \tag{34}$$

Taking into account (34), the function $\hat{\Phi}_{DP}$ can be written as

$$\hat{\Phi}_{DP} = \frac{1}{\sqrt{2}}\begin{pmatrix} f(\rho)\, i\sigma^3 \xi_{KN}(\theta) \\ g(\rho)\,\xi_{KN}(\theta) \end{pmatrix}. \tag{35}$$

Spinor $\xi_{KN}(\theta)$ is

$$\xi_{KN}(\theta) = \begin{pmatrix} S_-(\theta) \\ S_+(\theta) \end{pmatrix}, \tag{36}$$

$S_\mp(\theta)$ are spheroidal harmonics for a half-spin obeying the Chandrasekhar-Page angular equations. In absence of rotation, $\alpha_a = 0$ and

$$\xi_{KN}(\theta) = \xi(\theta) = \begin{pmatrix} {}_{-1/2}Y(\theta) \\ {}_{+1/2}Y(\theta) \end{pmatrix},$$

where ${}_{-1/2}Y(\theta)$, ${}_{+1/2}Y(\theta)$ are spherical harmonics for a half-spin (see, for example, [27], [28], [42]).

Spinor $\xi_{KN}(\theta)$ satisfies the equation

$$\left( i\sigma^2\left(\frac{\partial}{\partial\theta} + \frac{1}{2}\operatorname{ctg}\theta\right) + \sigma^1\left(-\alpha_a \varepsilon \sin\theta + \frac{m_\varphi}{\sin\theta}\right) - \sigma^3 \alpha_a \cos\theta \right)\xi_{KN}(\theta) = -\lambda \xi_{KN}(\theta). \tag{37}$$

From (37), the angular Chandrasekhar-Page equations for spheroidal harmonics $S_\mp(\theta)$ follow

$$\left(\frac{\partial}{\partial\theta} + \frac{1}{2}\operatorname{ctg}\theta\right)S_+ - \left(\alpha_a \varepsilon \sin\theta - \frac{m_\varphi}{\sin\theta}\right)S_+ = -\left(\lambda - \alpha_a\cos\theta\right)S_-,$$
$$\left(\frac{\partial}{\partial\theta} + \frac{1}{2}\operatorname{ctg}\theta\right)S_- + \left(\alpha_a \varepsilon \sin\theta - \frac{m_\varphi}{\sin\theta}\right)S_- = \left(\lambda + \alpha_a\cos\theta\right)S_+. \tag{38}$$

As opposed to Schwarzschild and Reissner-Nordström fields, the constant of separation $\lambda$ in (38) depends on $\varepsilon, \alpha_a, j, l, m_\varphi$.

Transformed Dirac equation (21) for function $\hat{\Psi}_{DP}$ will have the form

$$\left(i\gamma^3\right)M\left(R + A\right)M^{-1}\hat{\Psi}_{DP} = 0. \tag{39}$$

In (39), for conveniences, the transformed equation on the left is multiplied by matrix $i\gamma^3$. In addition, for comparison with the Dirac equation in Schwarzschild and Reissner-Nordström fields, we will use the next wave function in Eq. (39).

$$\Psi_{KN} = \frac{\hat{\Psi}_{DP}}{\rho\sqrt{f_{KN}}}. \tag{40}$$

Prior to writing Eq. (39), taking into account (40) in explicit form, we should note that when, using real radial wave functions $f(\rho), g(\rho)$, we must use positive values of $\Delta_{KN} \geq 0,\ f_{KN} \geq 0$. In this case, $\hat{\varepsilon}(\Delta_{KN}) = 1$. In presence of event horizons, the condition $\Delta_{KN} \geq 0$ excludes the spherical layer between outer and inner event horizons from the domain of wave functions.

As a result, the Dirac equation for $\Psi_{KN}$ (40) has the form

$$\begin{aligned}\left(\omega - \gamma^0\sqrt{f_{KN}}\right)\Psi_{KN} = &\left\{-i\gamma^0\gamma^3\left(f_{KN}\frac{\partial}{\partial\rho} + \frac{1}{\rho} - \frac{\alpha}{\rho^2}\right) + \frac{\sqrt{f_{KN}}}{\rho}\left[-i\gamma^0\gamma^5\alpha_a\cos\theta - \right.\right.\\ &\left.\left.-i\gamma^0\gamma^1\left(\frac{\partial}{\partial\theta} + \frac{1}{2}\mathrm{ctg}\theta\right) - \gamma^0\gamma^2\left(\varepsilon\alpha_a\sin\theta - \frac{m_\varphi}{\sin\theta}\right)\right]\right\}\Psi_{KN}.\end{aligned} \tag{41}$$

In (41), $\gamma^5 = \begin{pmatrix} 0 & 1 \\ 1 & 0 \end{pmatrix}$,

$$\omega = \varepsilon\left(1 + \frac{\alpha_a^2}{\rho^2}\right) - \frac{\alpha_a m_\varphi}{\rho^2} - \frac{\alpha_{em}}{\rho}. \tag{42}$$

Equation (41) is self-conjugate and, in absence of rotation $(\alpha_a = 0)$, coincides with the Dirac equation in the Reissner-Nordström field [28].

### 2.3. Separation of variables

The standard procedure of separation of variables is admissible for Eq. (41). First, (41) is written as a system of equations for upper and lower spinors of function $\Psi_{KN}$. Then, we use Eq. (37) for $\xi_{KN}(\theta)$. In the end, we obtain the equation for the real radial functions of $F_{KN}(\rho) = f(\rho)/\rho\sqrt{f_{KN}}$, $G_{KN}(\rho) = g(\rho)/\rho\sqrt{f_{KN}}$:

$$\begin{aligned}&\left(f_{KN}\frac{d}{d\rho} + \frac{1}{\rho} - \frac{\alpha}{\rho^2} + \frac{\lambda\sqrt{f_{KN}}}{\rho}\right)F_{KN}(\rho) - \left(\varepsilon\left(1 + \frac{\alpha_a^2}{\rho^2}\right) - \frac{\alpha_a m_\varphi}{\rho^2} - \frac{\alpha_{em}}{\rho} + \sqrt{f_{KN}}\right)G_{KN}(\rho) = 0,\\ &\left(f_{KN}\frac{d}{d\rho} + \frac{1}{\rho} - \frac{\alpha}{\rho^2} - \frac{\lambda\sqrt{f_{KN}}}{\rho}\right)G_{KN}(\rho) + \left(\varepsilon\left(1 + \frac{\alpha_a^2}{\rho^2}\right) - \frac{\alpha_a m_\varphi}{\rho^2} - \frac{\alpha_{em}}{\rho} - \sqrt{f_{KN}}\right)F_{KN}(\rho) = 0.\end{aligned} \tag{43}$$

Equations (43) are similar in their structure to equations for the Reissner-Nordström field. At $\alpha_a = 0$ (no rotation), Eqs. (43) coincide with the system of equations for the Reissner-Nordström field.

### 2.4. Asymptotics of radial wave functions

#### 2.4.1. Availability of event horizons $(\rho_+)_{KN}, (\rho_-)_{KN}$, $\alpha^2 > \alpha_a^2 + \alpha_Q^2$

It follows from Eqs. (43) that at $\rho \to \infty$ the leading terms of asymptotics are

$$F_{KN} = C_1\varphi_1(\rho)e^{-\sqrt{1-\varepsilon^2}\rho} + C_2\varphi_2(\rho)e^{\sqrt{1-\varepsilon^2}\rho},$$
$$G_{KN} = \frac{\sqrt{1-\varepsilon}}{\sqrt{1+\varepsilon}}\left(-C_1\varphi_1(\rho)e^{-\sqrt{1-\varepsilon^2}\rho} + C_2\varphi_2(\rho)e^{\sqrt{1-\varepsilon^2}\rho}\right). \tag{44}$$

In (44) $\varphi_1(\rho), \varphi_2(\rho)$ are power functions of $\rho$; $C_1, C_2$ are constants of integration.

When determining bound states of Dirac particles, it is necessary to use only exponentially decreasing solutions (44), i.e., in this case, $C_2 = 0$.

At $\rho \to \rho_+$ , let us present functions $F_{KN}(\rho), G_{KN}(\rho)$ as

$$F_{KN}\big|_{\rho\to(\rho_+)_{KN}} = \left|\rho-(\rho_+)_{KN}\right|^{s_+}\sum_{k=0}^{\infty} f_k^{(+)}\left|\rho-(\rho_+)_{KN}\right|^k,$$
$$G_{KN}\big|_{\rho\to(\rho_+)_{KN}} = \left|\rho-(\rho_+)_{KN}\right|^{s_+}\sum_{k=0}^{\infty} g_k^{(+)}\left|\rho-(\rho_+)_{KN}\right|^k. \tag{45}$$

The indicial equation for system (43) leads to the solution

$$s_+ = -\frac{1}{2} \pm i\frac{(\rho_+)_{KN}^2}{(\rho_+)_{KN}-(\rho_-)_{KN}}\left(\varepsilon\left(1+\frac{\alpha_a^2}{(\rho_+)_{KN}^2}\right) - \frac{\alpha_a m_\varphi}{(\rho_+)_{KN}^2} - \frac{\alpha_{em}}{(\rho_+)_{KN}}\right). \tag{46}$$

Formulas (45), (46) show that functions $F_{KN}(\rho), G_{KN}(\rho)$ are square nonintegrable on the outer event horizon of $\rho = (\rho_+)_{KN}$. The form of the oscillating part of function (45) for $\varepsilon \neq \dfrac{\alpha_a m_\varphi + \alpha_{em}(\rho_+)_{KN}}{\alpha_a^2 + (\rho_+)_{KN}^2}$ testifies to implementation of the mode of a particle "fall" to the outer event horizon [43], [44].

At $\rho \to (\rho_-)_{KN}$, let us present functions $F_{KN}(\rho), G_{KN}(\rho)$ as

$$F_{KN}\big|_{\rho\to(\rho_-)_{KN}} = \left|(\rho_-)_{KN}-\rho\right|^{s_-}\sum_{k=0}^{\infty} f_k^{(-)}\left|(\rho_-)_{KN}-\rho\right|^k,$$
$$G_{KN}\big|_{\rho\to(\rho_-)_{KN}} = \left|(\rho_-)_{KN}-\rho\right|^{s_-}\sum_{k=0}^{\infty} g_k^{(-)}\left|(\rho_-)_{KN}-\rho\right|^k. \tag{47}$$

In this case, the solution of the indicial equation for system (43) is

$$s_- = -\frac{1}{2} \pm i\frac{(\rho_-)_{KN}^2}{(\rho_+)_{KN}-(\rho_-)_{KN}}\left(\varepsilon\left(1+\frac{\alpha_a^2}{(\rho_-)_{KN}^2}\right) - \frac{\alpha_a m_\varphi}{(\rho_-)_{KN}^2} - \frac{\alpha_{em}}{(\rho_-)_{KN}}\right). \tag{48}$$

Expressions (47) and (48) show that $F_{KN}(\rho), G_{KN}(\rho)$ are square nonintegrable on the inner event horizon $\rho = (\rho_-)_{KN}$. The form of the oscillating parts of functions (47) for $\varepsilon \neq \dfrac{\alpha_a m_\varphi + \alpha_{em}(\rho_-)_{KN}}{\alpha_a^2 + (\rho_-)_{KN}^2}$ testifies to implementation of the mode of a particle “fall” to the inner event horizon [43], [44].

It is known [19] that at $\rho \to 0$, there are two square integrable solutions of the Dirac equation to the Kerr-Newman field. If we present

$$
\begin{aligned}
F_{KN}\big|_{\rho\to 0} &= \rho^s \sum_{k=0}^{\infty} f_k^{(0)} \rho^k, \\
G_{KN}\big|_{\rho\to 0} &= \rho^s \sum_{k=0}^{\infty} g_k^{(0)} \rho^k,
\end{aligned}
\tag{49}
$$

then the solutions of the indicial equation for system (43) are $s_1 = 0, s_2 = 1$ [20]. For both the solutions $s_1, s_2$, the function $F_{KN}, G_{KN}$ at $\rho \to 0$ are square integrable. The solution of this problem is proposed in [21] with the use of two-sheeted topology of the Kerr-Newman metric, allowing implementation beginning of the integration of equations (43) from $r \to -\infty$.

The boundary conditions for exponentially decreasing solutions at $r \to -\infty$ follow from (44):

$$
\begin{aligned}
F_{KN} &= C_2 \varphi_2(\rho) e^{\sqrt{1-\varepsilon^2}\rho}, \\
G_{KN} &= \frac{\sqrt{1-\varepsilon}}{\sqrt{1+\varepsilon}} C_2 \varphi_2(\rho) e^{\sqrt{1-\varepsilon^2}\rho}.
\end{aligned}
\tag{50}
$$

### 2.4.2. Extreme KN field $\left((\rho_+)_{KN} = (\rho_-)_{KN} = \alpha,\ \alpha^2 = \alpha_a^2 + \alpha_Q^2\right)$

At $\rho \to \infty$, the asymptotics (44) is valid. As well as in the previous section, we substitute the ambiguous asymptotics at $\rho \to 0$ by the asymptotics at $\rho \to -\infty$ (50). At $\rho \to \alpha$, the leading singularity of the effective potential to the left and to the right from the horizon and at $\varepsilon \neq \dfrac{\alpha_a m_\varphi + \alpha_{em}\alpha}{\alpha_a^2 + \alpha^2}$ has the form of (see (85) and Appendix)

$$
U_{eff}^F\left(\varepsilon \neq \frac{\alpha_a m_\varphi + \alpha_{em}\alpha}{\alpha_a^2 + \alpha^2}\right) = -\frac{\alpha^4}{2(\rho-\alpha)^4}\left[\varepsilon\left(1 + \frac{\alpha_a^2}{\alpha^2}\right) - \frac{m_\varphi \alpha_a}{\alpha^2} - \frac{\alpha_{em}}{\alpha}\right]^2, \tag{51}
$$

which testifies to implementation of the mode of a particle “fall” to the event horizon.

**2.4.3. Naked KN singularity** $\left(\alpha^2 < \alpha_a^2 + \alpha_Q^2\right); \rho \in (0, \infty)$

At $\rho \to \infty$, the asymptotics (44) is valid. As well as in sections 2.4.1, 2.4.2, we will use the asymptotics at $\rho \to -\infty$ (50) instead of the ambiguous asymptotics at $\rho \to 0$.

**2.5. Stationary solutions of Dirac equation**

**2.5.1. Nonregular stationary solutions of Dirac equation**

In presence of event horizons $(\rho_+)_{KN}, (\rho_-)_{KN}$, it follows from expressions (45) - (48) that there exist solutions, for which the oscillating part of the considered asymptotics is absent:

$$\varepsilon = \frac{\alpha_a m_\varphi + \alpha_{em} (\rho_\pm)_{KN}}{\alpha_a^2 + (\rho_\pm)_{KN}^2} \text{ - KN metric,} \tag{52}$$

$$\varepsilon = \frac{\alpha_a m_\varphi}{\alpha_a^2 + (\rho_\pm)_K^2} = \frac{\alpha_a m_\varphi}{2\alpha (\rho_\pm)_K} \text{ - Kerr metric (K), } \alpha_{em} = 0, \alpha_Q = 0. \tag{53}$$

In the absence of rotation $(\alpha_a = 0)$, expressions (52) - (53) coinside with the solutions for the Reissner-Nordström (RN) metric and Schwarzschild (S) metric, explored in detail in [27], [28]:

$$\varepsilon = \frac{\alpha_{em}}{(\rho_\pm)_{RN}} \text{ - RN metric,} \tag{54}$$

$$\varepsilon = 0 \text{ - S metric.} \tag{55}$$

For these solutions of the Dirac equation, there is no mode of a particle "fall" to the event horizons, however, solutions (52) and (53) are nonregular because of divergence of normalization integrals of asymptotics of wave functions (45) and (47) near the event horizons.

**2.5.2. Regular stationary solutions of the Dirac equation for the extreme KN field** $\left((\rho_+)_{KN} = (\rho_-)_{KN} = \alpha,\ \alpha^2 = \alpha_a^2 + \alpha_Q^2\right)$

In this case, the solutions (52) and (53) are

$$\varepsilon_{ext} = \frac{\alpha_a m_\varphi + \alpha_{em} \alpha}{\alpha_a^2 + \alpha^2} \text{ - KN metric,} \tag{56}$$

$$\varepsilon_{ext} = \frac{m_\varphi}{2\alpha} \text{ - K metric, } \alpha_{em} = 0, \alpha_Q = 0. \tag{57}$$

As opposed to (52) and (53), solutions (56) and (57) can be regular at definite ratios of initial parameters.

In the equations for real radial functions (43), we will denote

$$\omega(\rho)=\varepsilon\left(1+\frac{\alpha_a^2}{\rho^2}\right)-\frac{\alpha_a m_\varphi}{\rho^2}-\frac{\alpha_{em}}{\rho}. \tag{58}$$

For solutions (56) and (57)

$$\omega(\rho=\alpha,\varepsilon=\varepsilon_{ext})=0. \tag{59}$$

The first nonvanishing term in expansion $\omega(\rho,\varepsilon=\varepsilon_{ext})$ in power $|\rho-\alpha|$ is

$$\omega(\varepsilon=\varepsilon_{ext})\Big|_{\rho\to\alpha}=\Omega|\rho-\alpha|, \tag{60}$$

where

$$\Omega=-\frac{m_\varphi\alpha_a+\alpha_{em}\alpha}{\alpha^2+\alpha_a^2}\frac{2\alpha_a^2}{\alpha^3}+\frac{2m_\varphi\alpha_a}{\alpha^3}+\frac{\alpha_{em}}{\alpha^2}. \tag{61}$$

Taking into account that for the extreme KN field

$$f_{KN}^{ext}=\frac{(\rho-\alpha)^2}{\rho^2}, \tag{62}$$

Eqs. (43) at $\rho\to\alpha$ can be written as

$$\begin{aligned}&\left(\frac{|\rho-\alpha|}{\alpha^2}\frac{d}{d\rho}+\frac{1+\lambda}{\alpha^2}\right)F_{KN}-\left(\Omega+\frac{1}{\alpha}\right)G_{KN}=0,\\&\left(\frac{|\rho-\alpha|}{\alpha^2}\frac{d}{d\rho}+\frac{1-\lambda}{\alpha^2}\right)G_{KN}+\left(\Omega-\frac{1}{\alpha}\right)F_{KN}=0.\end{aligned} \tag{63}$$

If we present

$$\begin{aligned}F_{KN}\Big|_{\rho\to\alpha}&=|\rho-\alpha|^s\sum_{k=0}^{\infty}f_k^{ext}|\rho-\alpha|^k,\\G_{KN}\Big|_{\rho\to\alpha}&=|\rho-\alpha|^s\sum_{k=0}^{\infty}g_k^{ext}|\rho-\alpha|^k,\end{aligned} \tag{64}$$

then the solution of the fidicial equation for system (63) is

$$s_{1,2}=-1\pm\sqrt{\lambda^2+\alpha^2-\alpha^4\Omega^2}. \tag{65}$$

For square integrability of functions $F_{KN},G_{KN}$ in (63), it is necessary to retain solution (65) with the positive sign in front of the square root and, in addition, the following inequality should be fulfilled:

$$-2+2\sqrt{\lambda^2+\alpha^2-\alpha^4\Omega^2}>-1, \tag{66}$$

i.e.,

$$\lambda^2+\alpha^2-\alpha^4\Omega^2>1/4. \tag{67}$$

For the RN metric, $\alpha^4\Omega^2 = \alpha_{em}^2$, $\lambda^2 = \kappa^2$. In this case, the condition (67) coincides with the similar condition determined in [45].

### 2.5.3. Regular stationary solutions of Dirac equation for KN naked singularity $\left(\alpha^2 < \alpha_Q^2 + \alpha_a^2\right)$

When taking into account the two-sheeted structure of the KN solution and integrating the system of equations (43) from $\rho = -\infty$ to $\rho = +\infty$, the uncertainty in choosing the solution at $\rho = 0$ disappears and the discrete energy spectrum of fermions with square integrable wave functions will exist at definite values of physical parameters.

So, we validated the conclusions [8] - [10], [15], concerning the absence of stationary bound states of Dirac particles in the KN field in presence of event horizons. These conclusions also refer to Reissner-Nordström and Schwarzschild fields. For the extreme Kerr-Newman, Kerr fields, stationary bound states of Dirac half-spin particles exist for energies (56) and (57) provided condition (67) is met. Stationary solutions of the Dirac equation also exist for the case of naked KN singularity.

Now, let us turn to the second-order relativistic equation with the effective potential for the KN metric. By using this equation, we obtain the results qualitatively different from those in section 2.

## 3. Self-conjugate second-order equations for spinor wave functions of fermions in gravitational and electromagnetic Kerr-Neaman fields

In order to obtain second-order equations, it is necessary to perform the following [46]:

1. Squaring of Dirac equation.
2. Transition from bispinor to spinor wave functions in the second-order equation.
3. Nonunitary similarity transformation to ensure self-conjugacy of the second-order equation with spinor wave functions.

As a result, upon separation of angular and radial variables we can obtain the Schrödinger-type equation with the effective potential for transformed radial function $\psi_{KN}(\rho)$.

Such equations can be obtained from system (43).

Let us denote:

$$A_{KN} = -\frac{1}{f_{KN}}\left(\frac{1+\lambda\sqrt{f_{KN}}}{\rho} - \frac{\alpha}{\rho^2}\right), \tag{68}$$

$$B_{KN} = \frac{1}{f_{KN}}\left(\omega + \sqrt{f_{KN}}\right), \tag{69}$$

$$C_{KN} = -\frac{1}{f_{KN}}\left(\omega - \sqrt{f_{KN}}\right), \tag{70}$$

$$D_{KN} = -\frac{1}{f_{KN}}\left(\frac{1-\lambda\sqrt{f_{KN}}}{\rho} - \frac{\alpha}{\rho^2}\right). \tag{71}$$

In formulas (69) and (70), expression (42) is denoted as $\omega$.

Next, if we perform transformations

$$\psi_{KN}^F = g_F F_{KN}(\rho), \tag{72}$$

$$\psi_{KN}^G = g_G G_{KN}(\rho), \tag{73}$$

where

$$\begin{aligned} g_F(\rho) &= \exp\left(\frac{1}{2}\int^{\rho} A_F(\rho')\,d\rho'\right), \\ A_F(\rho) &= -\frac{1}{B}\frac{dB_{KN}}{d\rho} - A_{KN} - D_{KN}, \end{aligned} \tag{74}$$

$$\begin{aligned} g_G(\rho) &= \exp\left(\frac{1}{2}\int^{\rho} A_G(\rho')\,d\rho'\right), \\ A_G(\rho) &= -\frac{1}{C}\frac{dC_{KN}}{d\rho} - A_{KN} - D_{KN}, \end{aligned} \tag{75}$$

then, for $\psi_{KN}^F$ and $\psi_{KN}^G$, we will obtain the Schrödinger-type equations with effective potentials $U_{eff}^F(\rho), U_{eff}^G(\rho)$:

$$\frac{d^2\psi_{KN}^F}{d\rho^2} + 2\left(E_{Schr} - U_{eff}^F(\rho)\right)\psi_{KN}^F = 0, \tag{76}$$

$$\frac{d^2\psi_{KN}^G}{d\rho^2} + 2\left(E_{Schr} - U_{eff}^G(\rho)\right)\psi_{KN}^G = 0. \tag{77}$$

In (76) and (77)

$$E_{Schr} = \frac{1}{2}\left(\varepsilon^2 - 1\right). \tag{78}$$

In (76)

$$\begin{aligned} U_{eff}^F(\rho) = E_{Schr} &+ \frac{3}{8}\left(\frac{1}{B_{KN}}\frac{dB_{KN}}{d\rho}\right)^2 - \frac{1}{4}\frac{1}{B_{KN}}\frac{d^2B_{KN}}{d\rho^2} + \frac{1}{4}\frac{d}{d\rho}\left(A_{KN} - D_{KN}\right) - \\ &- \frac{1}{4}\frac{\left(A_{KN} - D_{KN}\right)}{B_{KN}}\frac{dB_{KN}}{d\rho} + \frac{1}{8}\left(A_{KN} - D_{KN}\right)^2 + \frac{1}{2}B_{KN}C_{KN}. \end{aligned} \tag{79}$$

The explicit form of potential (79) is presented in the Appendix.

In (77)

$$U^G_{eff}\left(\rho\right)=E_{Schr}+\frac{3}{8}\left(\frac{1}{C_{KN}}\frac{dC_{KN}}{d\rho}\right)^2-\frac{1}{4}\frac{1}{C_{KN}}\frac{d^2C_{KN}}{d\rho^2}-\frac{1}{4}\frac{d}{d\rho}\left(A_{KN}-D_{KN}\right)+$$
$$+\frac{1}{4}\frac{\left(A_{KN}-D_{KN}\right)}{C_{KN}}\frac{dC_{KN}}{d\rho}+\frac{1}{8}\left(A_{KN}-D_{KN}\right)^2+\frac{1}{2}B_{KN}C_{KN}. \quad (80)$$

In Eqs. (76) and (77), summand $E_{Schr}$ (78) is separated and simultaneously added to (79) and (80). This is done, on the one hand, for Eqs. (76) and (77) to take on Schrödinger-type forms and, on the other hand, to ensure classical asymptotics of effective potentials $U^F_{eff}$, $U^G_{eff}$ at $\rho\to\infty$.

The normalization integrals for wave functions $\psi^F_{KN}$ and $\psi^G_{KN}$ in (76) and (77) have the form of

$$N_F=\int^{\rho}\left(\psi^F_{KN}\left(\rho'\right)\right)^2 d\rho', \quad (81)$$

$$N_G=\int^{\rho}\left(\psi^G_{KN}\left(\rho'\right)\right)^2 d\rho'. \quad (82)$$

Equations (76) and (77) and effective potentials (79) and (80) transform to each other at $\varepsilon\to-\varepsilon,\lambda\to-\lambda,m_\varphi\to-m_\varphi,\alpha_{em}\to-\alpha_{em}$. Hence, it follows that Eqs/ (76) and (77) describe the motion of particles and antiparticles. In this paper, for particles, Eq. (76) is used for function $\psi^F_{KN}\left(\rho\right)$ with effective potential $U^F_{eff}$ (79).

The nonrelativistic limit of the Dirac equation with the lower spinor proportional to $G\left(\rho\right)$ and vanishing at zero momentum of particle $\left(\mathbf{p}=0\right)$ can serve as a basis for this. Similarly, the lower spinor with function $G\left(\rho\right)$ vanishes for the particle under the Foldy-Wouthuysen transformation with any value of momentum $\mathbf{p}$ [47] - [49]. On the other hand, the upper spinor of the bispinor wave function, proportional to $F\left(\rho\right)$, vanishes for the antiparticle at nonrelativistic limit $\mathbf{p}=0$ and under the Foldy-Wouthuysen transformation.

### 3.1. Singularities of effective potentials for stationary solutions $\varepsilon=\varepsilon_{KN}$

Here and below, for brevity, $\varepsilon_{KN}$ corresponds to stationary solutions (52) and (56).

**3.1.1.** At $\rho\to\pm\infty$

$$U^F_{eff}\left(\varepsilon_{KN}\right)\Big|_{\rho\to\pm\infty}=\varepsilon_{KN}\frac{\alpha_{em}}{\rho}+\left(1-2\varepsilon^2_{KN}\right)\frac{\alpha}{\rho}+\mathrm{O}\left(\frac{1}{\rho^2}\right). \quad (83)$$

**3.1.2.** In presence of two event horizons $(\rho_+)_{KN}, (\rho_-)_{KN}$

$$U_{eff}^F(\varepsilon_{KN})\Big|_{\rho\to(\rho_\pm)_{KN}} = -\frac{3}{32}\frac{1}{\left(\rho-(\rho_\pm)_{KN}\right)^2} + \mathrm{O}\left(\frac{1}{\left|\rho-(\rho_\pm)_{KN}\right|^{3/2}}\right). \qquad (84)$$

The asymptotics (84) are potential wells of $-K\big/\left(\rho-(\rho_\pm)_{KN}\right)^2$ with coefficient $K<1/8$, which testifies to possibility of existence of stationary bound states of quantum-mechnical particles near the event horizons (see, for example [43]).

**3.1.3.** For the extreme KN field $\left((\rho_+)_{KN} = (\rho_-)_{KN} = \alpha,\ \alpha^2 = \alpha_a^2 + \alpha_Q^2\right)$ at $\rho\to\alpha$ on the left and on the right from the event horizon:

$$U_{eff}^F(\varepsilon\neq\varepsilon_{KN})\Big|_{\rho\to\alpha} = -\frac{\left(\alpha_a^2+\alpha^2\right)^2\left(\varepsilon-\varepsilon_{KN}\right)^2}{2(\rho-\alpha)^4} + \mathrm{O}\left(\frac{1}{|\rho-\alpha|^3}\right), \qquad (85)$$

$$U_{eff}^F(\varepsilon=\varepsilon_{KN})\Big|_{\rho\to\alpha} = -\frac{1}{2(\rho-\alpha)^2}\left[\frac{1}{4}-\left(\lambda^2+\alpha^2-\alpha^4\Omega^2\right)\right] + \mathrm{O}\left(\frac{1}{|\rho-\alpha|}\right). \qquad (86)$$

From the asymptotics (86), for regular solutions (56), the condition of existence of a potential well and the condition of existence of stationary bound states $(K<1/8)$ therein can be written as

$$0<\lambda^2+\alpha^2-\alpha^4\Omega^2<1/4. \qquad (87)$$

**3.1.4.** At $\rho\to 0$

$$U_{eff}^F(\varepsilon_{KN})\Big|_{\rho\to 0} = \mathrm{const} + \mathrm{O}(\rho). \qquad (88)$$

As opposed to the Reissner-Nordström field $\left(U_{eff}^F(\varepsilon_{RN})\Big|_{\rho\to 0} = \dfrac{3}{8\rho^2}\right)$, effective KN potential (79) is regular at $\rho=0$.

**3.2. Impenetrable potential barriers**

The effective potential (79) at some values of $\rho=\rho_{cl}^i$ can have singularities of the following form:

$$U_{eff}^F\Big|_{\rho\to\rho_{cl}^i} \sim \frac{1}{\left(\omega+\sqrt{1-\dfrac{2\alpha}{\rho}+\dfrac{\alpha_a^2+\alpha_Q^2}{\rho^2}}\right)^2}, \qquad (89)$$

when the denominator in (89) becomes zero at one or several values of $\rho=\rho_{cl}^i$.

The equation for determining $\rho_{cl}^{i}$ has the form of

$$\varepsilon\left(1+\frac{\alpha_a^2}{\rho^2}\right)-\frac{m_\varphi\alpha_a}{\rho^2}-\frac{\alpha_{em}}{\rho}+\sqrt{1-\frac{2\alpha}{\rho}+\frac{\alpha_a^2+\alpha_Q^2}{\rho^2}}=0. \tag{90}$$

The singularities (89) can be contained in the second summand of effective potential (79) equal to $\frac{3}{8}\left(\frac{1}{B_{KN}}\frac{dB_{KN}}{d\rho}\right)^2$ (see the Appendix).

If the solution to Eq. (90) is available, potential (89) can be presented as

$$U_{eff}^{F}\Big|_{\rho\to\rho_{cl}^{i}}=\frac{3}{8\left(\rho-\rho_{cl}^{i}\right)^2}+\mathrm{O}\left(\frac{1}{\left|\rho-\rho_{cl}^{i}\right|}\right). \tag{91}$$

Such potential barriers are known to be impenetrable in terms of quantum mechanics [50]†.

Next, let us consider the conditions of emergence of singularities (91) in presence of two event horizons, for extreme KN fields and in case of the naked KN singularity.

**3.2.1. Presence of two event horizons $(\rho_+)_{KN},(\rho_-)_{KN},\alpha^2>\alpha_a^2+\alpha_Q^2$; domains of wave functions: $\rho\in\left[(\rho_+)_{KN},\infty\right),\rho\in\left(-\infty,(\rho_-)_{KN}\right]$**

In this section, we analyze the possibility of existence of impenetrable barriers of type (91) for stationary solutions of $\varepsilon_{KN}$ (52).

Equation (90) can be written as

$$\frac{\sqrt{\left(\rho-(\rho_+)_{KN}\right)\left(\rho-(\rho_-)_{KN}\right)}}{\rho}=-\varepsilon_{KN}\left(1+\frac{\alpha_a^2}{\rho^2}\right)+\frac{m_\varphi\alpha_a}{\rho^2}+\frac{\alpha_{em}}{\rho}. \tag{92}$$

The left and right parts of (92) are positive and vanish at $\rho=(\rho_\pm)_{KN}$. For the two domains of wave functions we can present (92) as

$$\rho\sqrt{\left(\rho-(\rho_+)_{KN}\right)\left(\rho-(\rho_-)_{KN}\right)}=\left(-\varepsilon_{KN}^{+}\rho-\frac{\alpha_a m_\varphi}{(\rho_+)_{KN}}+\frac{\alpha_a^2\varepsilon_{KN}^{+}}{(\rho_+)_{KN}}\right)\left(\rho-(\rho_+)_{KN}\right);$$
$$\rho\in\left[(\rho_+)_{KN},\infty\right), \tag{93}$$

$$\rho\sqrt{\left((\rho_+)_{KN}-\rho\right)\left((\rho_-)_{KN}-\rho\right)}=\left(\varepsilon_{KN}^{-}\rho+\frac{\alpha_a m_\varphi}{(\rho_-)_{KN}}-\frac{\alpha_a^2\varepsilon_{KN}^{-}}{(\rho_-)_{KN}}\right)\left((\rho_-)_{KN}-\rho\right);$$
$$\rho\in\left(-\infty,(\rho_-)_{KN}\right]. \tag{94}$$

---

† Note that the authors [50] used Schrödinger-type equation (76) without factor 2. In our notations, barrier $K/\left(\rho-\rho_{cl}^{i}\right)^2$ is impenetrable if $K\geq 3/8$.

Solutions $\rho = (\rho_+)_{KN}$ in (93) and $\rho = (\rho_-)_{KN}$ in (94) correspond to the singularities of effective potential $\left.\frac{3}{8}\left(\frac{1}{B_{KN}}\frac{dB_{KN}}{d\rho}\right)^2\right|_{\rho\to(\rho_\pm)_{KN}}$ on the event horizons. These singularities are already taken into account during the analysis in section 3.1.2.

Dividing both parts of (93) by $\sqrt{\rho-(\rho_+)_{KN}}$ and both parts of (94) by $\sqrt{(\rho_-)_{KN}-\rho}$, we will obtain equations for determining other singularities for $\rho \neq (\rho_+)_{KN}$ and $\rho \neq (\rho_-)_{KN}$, respectively:

$$\rho\sqrt{\rho-(\rho_-)_{KN}} = \left(-\varepsilon^+_{KN}\rho - \frac{\alpha_a m_\varphi}{(\rho_+)_{KN}} + \frac{\alpha_a^2 \varepsilon^+_{KN}}{(\rho_+)_{KN}}\right)\sqrt{\rho-(\rho_+)_{KN}};$$
$$\rho \in \left((\rho_+)_{KN}, \infty\right), \tag{95}$$

$$\varepsilon^+_{KN} = \frac{\alpha_a m_\varphi + \alpha_{em}(\rho_+)_{KN}}{\alpha_a^2 + (\rho_+)^2_{KN}}, \tag{96}$$

$$\rho\sqrt{(\rho_+)_{KN}-\rho} = \left(\varepsilon^-_{KN}\rho + \frac{\alpha_a m_\varphi}{(\rho_-)_{KN}} - \frac{\alpha_a^2 \varepsilon^-_{KN}}{(\rho_-)_{KN}}\right)\sqrt{(\rho_-)_{KN}-\rho};$$
$$\rho \in \left(0, (\rho_-)_{KN}\right), \tag{97}$$

$$\varepsilon^-_{KN} = \frac{\alpha_a m_\varphi + \alpha_{em}(\rho_-)_{KN}}{\alpha_a^2 + (\rho_-)^2_{KN}}. \tag{98}$$

Next, we will consider the following three variants for analysis:

1. The Kerr field: $\alpha_Q = 0,\ \alpha_{em} = 0,\ (\rho_\pm)_K = \alpha \pm \sqrt{\alpha^2 - \alpha_a^2}$,

$\varepsilon_K = \frac{\alpha_a m_\varphi}{2\alpha(\rho_\pm)_K},\ m_\varphi > 0,\ 0 < \varepsilon_K < 1.$

2. The uncharged half-spin particle in the Kerr-Newman field: $\alpha_{em} = 0, \alpha_Q \neq 0$,

$(\rho_\pm)_{KN} = \alpha \pm \sqrt{\alpha^2 - \alpha_Q^2 - \alpha_a^2},\ \varepsilon_{KN} = \frac{\alpha_a m_\varphi}{\alpha_a^2 + (\rho_\pm)^2_{KN}}, m_\varphi > 0,\ 0 < \varepsilon_{KN} < 1.$

The inequalities in variants 1, 2 for $m_\varphi, \varepsilon_K, \varepsilon_{KN}$ will be discussed in section 5.

3. The charged half-spin partilce in the Kerr-Newman field: $\alpha_{em} \neq 0, \alpha_Q \neq 0$,

$(\rho_\pm)_{KN} = \alpha \pm \sqrt{\alpha^2 - \alpha_a^2 - \alpha_Q^2},\ \varepsilon_{KN} = \frac{\alpha_a m_\varphi + \alpha_{em}(\rho_\pm)_{KN}}{\alpha_a^2 + (\rho_\pm)^2_{KN}}.$

There is no solution of Eq. (95) for domain $\rho > (\rho_+)_{KN}$ and for the first two variants. On the other hand, for the first two variants and for domain $\rho < (\rho_-)_{KN}$ , there exists the single solution (97) of $\rho = \rho_{cl}$ , at which the impenetrable potential barrier of type (91) emerges.

For the third variant and for both the domains of $\rho > (\rho_+)_{KN}$ , $\rho < (\rho_-)_{KN}$ , the algebrtaic analysis of admissible parameters, satisfying Eqs. (95) and (97) is difficult because of possible implementation of different signs of $\varepsilon, m_\varphi, \alpha_{em}$ . In this case, the possibility of existence of the impenetrable potential barrier at specified initial parameters should be determined either by the analysis of the effective potential or by the solution of Eqs. (95) and (97).

**3.2.2. Extreme KN field** $\left(\alpha^2 = \alpha_a^2 + \alpha_Q^2,\ (\rho_+)_{KN} = (\rho_-)_{KN} = \alpha\right)$**; domain of wave functions:** $\rho \in [\alpha, \infty), \rho \in (-\infty, \alpha]$

In case of the extreme KN field for both the domains of wave functions, Eq. (90) can be written as

$$\rho(\rho - \alpha) = \left(-\varepsilon_{KN}^{ext}\rho - \frac{\alpha_a m_\varphi}{\alpha} + \frac{\alpha_a^2}{\alpha}\varepsilon_{KN}^{ext}\right)(\rho - \alpha), \tag{99}$$

$$\varepsilon_{KN}^{ext} = \frac{\alpha_a m_\varphi + \alpha_{em}\alpha}{\alpha_a^2 + \alpha^2}. \tag{100}$$

The solution of $\rho = \alpha$ in Eq. (99) corresponds to the singularity of effective potential $\left.\frac{3}{8}\left(\frac{1}{B_{KN}}\frac{dB_{KN}}{d\rho}\right)^2\right|_{\rho\to\alpha}$ on the event horizon. This singularity is taken into account in the analysis in section 3.1.3.

Next, we will examine Eq. (99) for $\rho \neq \alpha$ . As in section 3.2.1, let us consider the following three variants for the analysis:

1. The Kerr field: $\alpha_Q = 0,\ \alpha_{em} = 0,\ \alpha = \alpha_a,\ \varepsilon_K = m_\varphi / 2\alpha,\ m_\varphi > 0;\ 0 < \varepsilon_K^{ext} < 1.$

2. The uncharged half-spin particle in the Kerr-Newman field: $\alpha_{em} = 0, \alpha_Q \neq 0,$

$\alpha^2 = \alpha_a^2 + \alpha_Q^2,\ \varepsilon_{KN}^{ext} = \frac{\alpha_a m_\varphi}{\alpha_a^2 + \alpha^2},\ m_\varphi > 0;\ 0 < \varepsilon_{KN}^{ext} < 1.$

The inequality in variants 1.2. for $m_\varphi,\ \varepsilon_K,\ \varepsilon_{KN}$ will be discussed in section 5.

3. The charged half-spin particle in the Kerr-Newman field: $\alpha_{em} \neq 0, \alpha_Q \neq 0,$

$$\alpha^2 = \alpha_a^2 + \alpha_Q^2,\ \varepsilon_{KN}^{ext} = \frac{\alpha_a m_\varphi + \alpha_{em}\alpha}{\alpha_a^2 + \alpha^2}.$$

There are no solutions of Eq. (99) for domain $\rho > \alpha$ and for the first two variants. On the other hand, for the first two variants and for domain $\rho < \alpha$, there exists the single solution of $\rho = \rho_{cl}$, at which the impenetrable potential barrier of type (91) emerges.

For the third variant and for both the domains of $\rho > \alpha,\ \rho < \alpha$, the algebrtaic analysis is difficult. The possibility of existence of the impenetrable potential barrier at specified initial parameters should be determined by solution of Eq. (99).

**3.2.3. KN naked singularity $\left(\alpha_a^2 + \alpha_Q^2 > \alpha^2\right)$; domain of wave function: $\rho \in (-\infty, \infty)$**

In case of KN naked singularity, it is necessary to solve the fourth-order equation (90) to determine presence or absence of impenetrable potential barriers. Because of cumbersome analytical solutions, it is reasonable to solve Eq. (90) at specified values of initial parameters: $\alpha, \alpha_Q, \alpha_a, \alpha_{em}, m_\varphi, \varepsilon$. The second method involves computer investigation of the singularities of effective potential (79) for some designated domain of initial values of the parameters.

**3.3. Asymptotics of functions $\psi_{KN}^F(\rho,\varepsilon)$**

**3.3.1.** At $\rho \to \pm\infty$ $\quad g_F(\rho) \to \rho$ (see (74)) and

$$\psi_{KN}^F\Big|_{\rho\to\pm\infty} = \rho F_{KN}\Big|_{\rho\to\pm\infty}. \tag{101}$$

For the finite motion of half-spin particles, taking into account (44),

$$\psi_{KN}^F\Big|_{\rho\to\infty} = C_1\varphi_1(\rho)\rho e^{-\sqrt{1-\varepsilon^2}\rho}, \tag{102}$$

$$\psi_{KN}^F\Big|_{\rho\to-\infty} = C_2\varphi_2(\rho)\rho e^{\sqrt{1-\varepsilon^2}\rho} \tag{103}$$

**3.3.2.** In presence of two event horizons, let us present function $\psi_{KN}^F(\rho, \varepsilon_{KN})$

at $\rho \to (\rho_+)_{KN}$

$$\psi_{KN}^F(\varepsilon_{KN})\Big|_{\rho\to(\rho_+)_{KN}} = \left|\rho - (\rho_+)_{KN}\right|^s \sum_{k=0}^{\infty} \chi_k^{(+)} \left|\rho - (\rho_+)_{KN}\right|^k, \tag{104}$$

and at $\rho \to (\rho_-)_{KN}$

$$\psi_{KN}^F(\varepsilon_{KN})\Big|_{\rho\to(\rho_-)_{KN}} = \left|(\rho_-)_{KN} - \rho\right|^s \sum_{k=0}^{\infty} \chi_k^{(-)} \left|(\rho_-)_{KN} - \rho\right|^k. \tag{105}$$

The following indicial equation follows from Eq. (76), taking into account (104), (105), (84):

$$s\left(s-1\right)+3/16=0 \tag{106}$$

with the solutions of $s_1=3/4, s_2=1/4$.

Both solutions lead to regular square-integrable solutions for wave function $\psi_{KN}^F\left(\rho,\varepsilon_{KN}\right)$. For an unambiguous choice of the solution, let us turn to asymptotics (45) - (48) for Dirac radial functions $F_{KN}\left(\rho\right)$ at $\varepsilon=\varepsilon_{KN}$ and to transformation (72), (74).

At $\rho\to\left(\rho_\pm\right)_{KN}$, transformation $g_F\left(\rho\right)\to\left|\rho-\left(\rho_\pm\right)_{KN}\right|^{3/4}$ and as a result

$$\psi_{KN}^F\left(\varepsilon_{KN}\right)\Big|_{\rho\to\left(\rho_\pm\right)_{KN}}=\left[g_F\left(\rho\right)F_{KN}\left(\rho\right)\right]\Big|_{\rho\to\left(\rho_\pm\right)_{KN}}=C_3\left|\rho-\left(\rho_\pm\right)_{KN}\right|^{1/4}. \tag{107}$$

The asymptotics (107) correspond to the solution of indicial equation (106) $s_2=1/4$. For the solution of $s_1=3/4$, there is no solution to Dirac equation $F\left(\rho\right)\Big|_{\rho\to\left(\rho_\pm\right)_{KN}}=\text{const}$, and in this case transformation (72) does not exist. Below, we will use solutions of the second order equation (76) with asymptotics (107) as eigenfunctions of stationary bound states of fermions with eigen values of $\varepsilon_{KN}$. These solutions are square integrable in the neighborhood of event horizons. Let us note that wave functions (107) on event horizons $\left(\rho_+\right)_{KN},\left(\rho_-\right)_{KN}$ are zero.

**3.3.3.** In case of the extreme KN field, let us present the asympotics of function $\psi_{KN}^F\left(\rho,\varepsilon_{KN}^{ext}\right)$ as

$$\psi_{KN}^F\left(\rho,\varepsilon_{KN}^{ext}\right)\Big|_{\rho\to\alpha}=\left|\rho-\alpha\right|^s\sum_{k=0}^{\infty}\chi_k^{ext}\left|\rho-\alpha\right|^k. \tag{108}$$

We can write the following indicial equation from Eq. (76) taking into account (108) and (86):

$$s\left(s-1\right)+\frac{1}{4}-\lambda^2-\alpha^2+\alpha^4\Omega^2=0 \tag{109}$$

with the solutions

$$s_{1,2}=\frac{1}{2}\pm\sqrt{\lambda^2+\alpha^2-\alpha^4\Omega^2}. \tag{110}$$

Both the solutions lead to square-integrable functions $\psi_{KN}^F\left(\varepsilon_{KN}^{ext}\right)$, provided the following conditions are met:

- for $s_1=1/2+\sqrt{\lambda^2+\alpha^2-\alpha^4\Omega^2}$ , the following inequality must be sutisfied:

$$\lambda^2+\alpha^2-\alpha^4\Omega^2>0, \tag{111}$$

- for $s_2=1/2-\sqrt{\lambda^2+\alpha^2-\alpha^4\Omega^2}$ , the following inequality must be sutisfied:

$$1>\lambda^2+\alpha^2-\alpha^4\Omega^2>0. \tag{112}$$

The condition of existence of potential well (87) essentially limits inequalities (111) and (112), and both the solutions of $s_{1,2}$ should sutisfy the inequality

$$1/4 > \lambda^2 + \alpha^2 - \alpha^4\Omega^2 > 0. \tag{113}$$

At $\rho \to \alpha$, transformation (74) $g_F(\rho) \to |\rho - \alpha|^{3/2}$. Both solutions (65) for asymptotics $F_{KN}^{ext}\big|_{\rho\to\alpha}$ (64) are transformed to appropriate solutions (110) for asymptotics of $\psi_{KN}^F(\varepsilon_{KN}^{ext})$ (108).

Thus, it is impossible to formulate the boundary problem of existence of eigenfunctions and eigenvalues for second-order equation (76) because of existence of two regular asymptotics in the neighborhood of the event horizon of $\rho = \alpha$.

Moreover, the authors failed to find, analytically and numerically, the domain of physical parameters where inequality (113) is valid.

This problem is unambiguously resoved for the Dirac equation. The single solution, physically admissible for the wave function corresponds to one of the solutions of (65)

$$s_1 = -1 + \sqrt{\lambda^2 + \alpha^2 - \alpha^4\Omega^2}. \tag{114}$$

The condition of square integrability of Dirac radial functions (66) is

$$\lambda^2 + \alpha^2 - \alpha^4\Omega^2 > 1/4. \tag{115}$$

In this case, the energy of stationary bound state $\varepsilon_{KN}^{ext}$ is

$$\varepsilon_{KN}^{ext} = \frac{\alpha_a m_\varphi + \alpha_{em}\alpha}{\alpha_a^2 + \alpha^2}. \tag{116}$$

For the Kerr metric $(\alpha_{em} = 0,\ \alpha_Q = 0)$, expressions (115) and (116) are

$$\lambda^2 + \alpha^2 - m_\varphi^2 > \frac{1}{4},\ \varepsilon_K^{ext} = \frac{m_\varphi}{2\alpha}, \tag{117}$$

The solution of $\varepsilon_K^{ext}$ was earlier obtained in [51].

**3.3.4.** Let us consider the asymptotics of wave function $\psi_{KN}^F\big|_{\rho\to\rho_{cl}^i}$ near the impenetrable barriers (91). The indicial equation for (76) taking into account asymptotics (91) has the form

$$s(s-1) - 3/4 = 0. \tag{118}$$

Solutions of (118) are $s_1 = 3/2, s_2 = -1/2$. The second solution corresponds to unnormalized wave function $\psi_{KN}^F$ and therefore it is physically inadmissible. As the result,

$$\psi_{KN}^F\big|_{\rho\to\rho_{cl}^i} = C_4\left(\left|\rho - \rho_{cl}^i\right|\right)^{3/2}. \tag{119}$$

The transformation (72), (74) in the neighborhood of $\rho_{cl}^{i}$ is singular, $g_F(\rho) \to \left|\rho - \rho_{cl}^{i}\right|^{-1/2}$. It follows from (72) that

$$F_{KN}(\rho)\Big|_{\rho\to\rho_{cl}^{i}} \approx \left|\rho - \rho_{cl}^{i}\right|^{2}. \tag{120}$$

It follows from the numerical calculations that the relation (120) is really implemented in the first equation (43) for the case when the expression in the brackets in front of function $G_{KN}(\rho)$ vanishes (this corresponds to Eq. (90) for determining $\rho_{cl}^{i}$).

## 4. Method of phase functions

Numerical solutions of Schrödinger-type second-order equations (76) are convenient to implement by using phase functions $\Phi(\rho), \Phi(\theta)$.

Let us apply the Prüfer transformation [30] - [33] to Eq. (76) with effective potential (79):

$$\begin{aligned} \psi_{KN}^{F}(\rho) &= P(\rho)\sin\Phi(\rho), \\ \frac{d\psi_{KN}^{F}(\rho)}{d\rho} &= P(\rho)\cos\Phi(\rho). \end{aligned} \tag{121}$$

Then,

$$\psi_{KN}^{F}(\rho) \Big/ \frac{d\psi_{KN}^{F}(\rho)}{d\rho} = \mathrm{tg}\Phi(\rho) \tag{122}$$

and Eq. (76) can be written as the system of first-order nonlinear differential equations

$$\frac{d\Phi}{d\rho} = \cos^2\Phi + 2\left(E_{Schr} - U_{eff}^{F}\right)\sin^2\Phi, \tag{123}$$

$$\frac{d\ln P}{d\rho} = \left(1 - 2\left(E_{Schr} - U_{eff}^{F}\right)\right)\sin\Phi\cos\Phi. \tag{124}$$

Let us note that Eq. (124) should be solved after determining eigenvalues $\varepsilon_n$ and eigenfunctions $\Phi_n(\rho)$ from Eq. (123).

In the equations, effective potential $U_{eff}^{F}$ depends on separation constant $\lambda$. For Kerr and Kerr-Newman fields, as opposed to Schwarschild, Reissner-Nordström fields , $\lambda$ depends on $\varepsilon, \alpha_a, j, l, m_\varphi$ (see angular Chandrasekhar-Page equations (38) for spheroidal harmonics $S_\pm(\theta)$). When solving equations (123) and (124), it is necessary initially to determine values $\lambda$ versus the initial parameters.

Let us apply the Prüfer transformation to Eqs. (38).

$$S_{-}(\theta) = S(\theta)\sin\Phi(\theta),$$
$$S_{+}(\theta) = S(\theta)\cos\Phi(\theta), \tag{125}$$

where

$$S_{-}(\theta)/S_{+}(\theta) = \mathrm{tg}\Phi(\theta), \tag{126}$$

$$S(\theta) = \left(S_{-}^{2}(\theta) + S_{+}^{2}(\theta)\right)^{1/2}. \tag{127}$$

Then, angular Chandrasekhar-Page equations (38) can be written as

$$\frac{d\Phi(\theta)}{d\theta} = \lambda + \alpha_a \cos\theta \cos\left(2\Phi(\theta)\right) + \left(\frac{m_\varphi}{\sin\theta} - \alpha_a \varepsilon \sin\theta\right)\sin\left(2\Phi(\theta)\right), \tag{128}$$

$$\frac{d\ln S}{d\theta} = -\frac{1}{2}\mathrm{ctg}\theta + \left(\alpha_a \varepsilon \sin\theta - \frac{m_\varphi}{\sin\theta}\right)\cos\left(2\Phi(\theta)\right) + \alpha_a \cos\theta \sin\left(2\Phi(\theta)\right). \tag{129}$$

The boundary conditions have the form

$$\text{for } m_\varphi < 0 \quad \Phi(0) = k\pi, \Phi(\pi) = \frac{\pi}{2} + k\pi, \tag{130}$$

$$\text{for } m_\varphi > 0 \quad \Phi(0) = \frac{\pi}{2} + k\pi, \Phi(\pi) = k\pi. \tag{131}$$

In (130) and (131), $k = 0, \pm 1, \pm 2, ...$

$$\mathrm{tg}\Phi\left(\frac{\pi}{2}\right) = P_{par}(-1)^{j+m_\varphi} = \pm 1, \tag{132}$$

where $P_{par} = 2(l - j)$ is parity.

Earlier, V.P.Neznamov and I.I. Safronov [52] implemented numerical solutions of Eqs. (128) and (129). The solution results fully agree with the results of [42], obtained using another numerical approach.

In this paper, we will first determine $\lambda(\varepsilon, \alpha_a, l, j, m_\varphi)$ and $\Phi(\theta)$ from Eq. (128). Then, we determine the spectrum of $\varepsilon_n$ and $\Phi_n(\rho)$ from (123). We determine $P_n(\rho)$ from Eq. (124) and, if necessary, determine $S_n(\theta)$ from Eq. (129). Thereafter, we determine radial eigenfunction ${}_n\psi_{KN}^{F}(\rho)$ in compliance with (121) and, if necessary, the full wave function in compliance with (125) and (35), (36), (40):

$${}_n\psi_{KN}^{F}(\rho, \theta, \varphi) = {}_n\psi_{KN}^{F}(\rho)\, i\sigma^3 \xi_{KN}(\theta). \tag{133}$$

### 4.1. Asymptotics of functions $\Phi(\rho), P(\rho)$

**4.1.1.** For the bound states at $\rho \to \infty$, taking into account (102) and (122)

$$\begin{aligned} \mathrm{tg}\Phi\big|_{\rho\to\infty} &= -\frac{1}{\sqrt{1-\varepsilon^2}}, \\ \Phi\big|_{\rho\to\infty} &= -\mathrm{arctg}\frac{1}{\sqrt{1-\varepsilon^2}} + k\pi. \end{aligned} \tag{134}$$

For the exponentially growing solutions in asymptotics (44), $C_1 = 0,\ C_2 \neq 0$, and , taking into account (103) and (122)

$$\begin{aligned} \mathrm{tg}\Phi\big|_{\rho\to\infty} &= \frac{1}{\sqrt{1-\varepsilon^2}}, \\ \Phi\big|_{\rho\to\infty} &= \mathrm{arctg}\frac{1}{\sqrt{1-\varepsilon^2}} + k\pi. \end{aligned} \tag{135}$$

For bound states at $\rho \to -\infty$ , taking into account (103) and (122),

$$\begin{aligned} \mathrm{tg}\Phi\big|_{\rho\to-\infty} &= \frac{1}{\sqrt{1-\varepsilon^2}}, \\ \Phi\big|_{\rho\to-\infty} &= \mathrm{arctg}\frac{1}{\sqrt{1-\varepsilon^2}} + k\pi. \end{aligned} \tag{136}$$

For the exponentially growing solutions,

$$\begin{aligned} \mathrm{tg}\Phi\big|_{\rho\to-\infty} &= -\frac{1}{\sqrt{1-\varepsilon^2}}, \\ \Phi\big|_{\rho\to-\infty} &= -\mathrm{arctg}\frac{1}{\sqrt{1-\varepsilon^2}} + k\pi. \end{aligned} \tag{137}$$

In (134) - (137) $k = 0, \pm 1, \pm 2, \ldots$.

**4.1.2.** In presence of two event horizons, let, at $\rho \to (\rho_+)_{KN}$,

$$\Phi\big|_{\rho\to(\rho_+)_{KN}} = k\pi + A\left|\rho - (\rho_+)_{KN}\right|. \tag{138}$$

Then, $\sin\Phi\big|_{\rho\to(\rho_+)_{KN}} \simeq \pm A\left|\rho-(\rho_+)_{KN}\right|;\ \cos\Phi\big|_{\rho\to(\rho_+)_{KN}} \simeq \pm 1$.

From the consistency of (138) with Eq. (123), taking into account the leading singularity $U_{eff}^F(\varepsilon_{KN}) = -\frac{3}{32}\frac{1}{\left(\rho-(\rho_+)_{KN}\right)^2}$ (see (84)), we obtain

$$1 + \frac{3}{16}A^2 = A \tag{139}$$

with the solutions $A_1 = 4, A_2 = 4/3$.

Next, we integrate Eq. (124) at $\rho \to (\rho_+)_{KN}$, taking into account the leading singularity of effective potential (84). As the result,

$$P\big|_{\rho\to(\rho_+)_{KN}} = C_5 \begin{cases} \left|\rho-(\rho_+)_{KN}\right|^{3/4}, A_1 = 4 \\ \left|\rho-(\rho_+)_{KN}\right|^{-1/4}, A_2 = 4/3 \end{cases} \tag{140}$$

$$\psi^F_{KN}\left(\varepsilon^+_{KN}\right)\Big|_{\rho\to(\rho_+)_{KN}} = C_5 \begin{cases} 4\left|\rho-(\rho_+)_{KN}\right|^{1/4}, A_1 = 4 \\ 4/3\left|\rho-(\rho_+)_{KN}\right|^{3/4}, A_2 = 4/3 \end{cases} \tag{141}$$

Comparison with (107) shows that solutions (138), (140), and (141) with the values of $A_1 = 4$ and $C_3 = 4C_5$ are admissible for our consideration.

Likewise, at $\rho \to (\rho_-)_{KN}$,

$$\Phi\big|_{\rho\to(\rho_-)_{KN}} = -4\left|(\rho_-)_{KN} - \rho\right| + k\pi, \tag{142}$$

$$P\big|_{\rho\to(\rho_-)_{KN}} = -\frac{C_3}{4}\left|(\rho_-)_{KN} - \rho\right|^{-3/4}, \tag{143}$$

$$\psi^F_{KN}\left(\varepsilon_{KN}\right)\Big|_{\rho\to(\rho_-)_{KN}} = C_3\left|(\rho_-)_{KN} - \rho\right|^{1/4}. \tag{144}$$

**4.1.3.** By analogy with section 4.1.2, asymptotics $\Phi(\rho), P(\rho)$ in the neighborhood of impenetrable barriers (91), taking into account (119) , are

$$\Phi\big|_{\rho\to\rho^i_{cl}} = \frac{2}{3}\left|\rho-\rho^i_{cl}\right| + k\pi, \tag{145}$$

$$P\big|_{\rho\to\rho^i_{cl}} = C_6\left(\left|\rho-\rho^i_{cl}\right|\right)^{1/2}, \tag{146}$$

$$\psi^F_{KN}\Big|_{\rho\to\rho^i_{cl}} = C_4\left(\left|\rho-\rho^i_{cl}\right|\right)^{3/2}. \tag{147}$$

## 5. Stationary bound states of fermions in the Kerr-Newman field

### 5.1. Numerical solutions of equations for phase functions. Common properties of phase functions

Below, we will concentrate on the numerical solution method for Eqs. (123) and (124) for radial phase functions $\Phi(\rho), P(\rho)$. The similar solution method for Eqs. (128), (129) for angular phase functions $\Phi(\theta), P(\theta)$ is described in detail in [52].

For the allowed set of values of $-1 < \varepsilon < 1$, the Cauchy problem is numerically solved with the specified initial condition. For the solution of the Cauchy problem, we use the fifth-order Runge-Kutta explicit method with step control (the Ehle scheme of the Radau II A three-stage method [53]).

Having determined spectrum $\varepsilon_n$ and eigenfunctions $\Phi_n(\rho)$ by solution (123) and by integrating Eq. (124), we can determine function $P_n(\rho)$ and wave functions ${}_n\psi^F_{KN}(\rho)$, taking into account (121) . Next, we can determine the probability density of fermions detection in the state with $\varepsilon_n$ at distance $\rho$ in spherical layer $d\rho$,

$$w(\rho) = P_n^2(\rho)\sin^2\Phi_n(\rho) \tag{148}$$

and probability of detection of bound fermions within the range of $[\rho_0, \rho]$

$$W(\rho) = \int_{\rho_0}^{\rho} P_n^2(\rho')\sin^2\Phi_n(\rho')\,d\rho'. \tag{149}$$

In presence of two event horizons, the energy of bound states is determined by equalities (52) and (53). In this case, only eigenfunctions ${}_n\psi^F_{KN}(\rho)$ (121), probability densities (148), and integral probabilities (149) are determined numerically.

Wave functions ${}_n\psi^F_{KN}(\rho)$ should satisfy asymptotics (102), (103), and (107), depending on the domain.

Existence of singular points $\rho = \pm\infty,\ \rho = \rho^i_{cl},\ \rho = (\rho_\pm)_{KN}$ should be taken into account while solving the equation. The numerical calculations with good reproducibility of the results began or ended in the neighborhood of irregular singular points $\rho = (\rho_\pm)_{KN}$ , $\rho = \rho_{cl}$ with $\Delta\rho_{irr} = 10^{-8}$. The choice of the maximal value of $\rho_{\max}$ in the calculations with simulation of $\rho \to \pm\infty$ was determined by meeting conditions (134) - (137) with the specified accuracy of $10^{-7}$.

Below, we will use function $\Phi(\varepsilon, \rho_{\max}) = \Phi(\varepsilon)\big|_{\rho=\rho_{\max}}$ for the case of KN naked singularity while determining the spectrum of $\varepsilon_n$. Here $\rho_{\max}$ is the maximal distance in numerical calculations. As a rule, the value $\rho_{\max} = 10^7$ ensures good convergence of the results.

The numerical calculations revealed the presence of the following essential properties of function $\Phi(\varepsilon, \rho_{\max})$ (the similar properties of function $\Phi$ for simplier potentials independent of $\varepsilon$ were rigorously proved in [31] - [33]):

1. Function $\Phi(\varepsilon, \rho_{\max})$ is monotonous when $\varepsilon$ changes.

2. In the case of existence of bound states with $-1<\varepsilon<1$, the behavior of $\Phi(\varepsilon,\rho_{\max})$ is stepwise. When the eigenvalue of $\varepsilon_n$ is achieved, function $\Phi(\varepsilon,\rho_{\max})$ changes stepwise by $\pi$:

$$\left[\Phi(\varepsilon_0-\Delta\varepsilon,\rho_{\max})-\Phi(\varepsilon_n+\Delta\varepsilon,\rho_{\max})\right]\Big|_{\Delta\varepsilon\to 0}=-n\pi. \qquad (150)$$

3. In the case of absence of bound states, the variation in function $\Phi(\varepsilon,\rho_{\max})$ within the entire range of $-1<\varepsilon<1$ is lower than $\pi$.

**5.2. Availability of two event horizons $(\rho_+)_{KN}, (\rho_-)_{KN}; \alpha^2>\alpha_a^2+\alpha_Q^2$. Domain of wave functions $\rho\in\left[(\rho_+)_{KN},\infty\right)$**

In this case, solution (52) exists:

$$\varepsilon_{KN}=\frac{\alpha_a m_\varphi+\alpha_{em}(\rho_+)_{KN}}{\alpha_a^2+(\rho_+)_{KN}^2}. \qquad (151)$$

For bound states, $-1<\varepsilon_{KN}<1$, therefore,

$$\left|\alpha_a m_\varphi+\alpha_{em}(\rho_+)_{KN}\right|<\alpha_a^2+(\rho_+)_{KN}^2. \qquad (152)$$

Solution (151) includes the state with like $(\alpha_{em}>0)$ and opposite $(\alpha_{em}<0)$ charges of the RN field source and the fermion as well as the states with uncharged fermions.

While determining wave functions with known eigenvalue (151), Eq. (123) was integrated from "right to left", from $\rho=\rho_{\max}$ with boundary condition (134) to $\rho=(\rho_+)_{KN}$ with asymptotics (138) and with the solution of Eq. (139) $A_1=4$. The behavior of the integral curves of the equation (123) near irregular singular point $\rho=(\rho_+)_{KN}$ is similar to that considered earlier for the RN field [28]. It is associated with the similar behavior of the effective potentials at $\rho\to(\rho_+)_{KN}$ and at $\rho\to(\rho_+)_{RN}$ (see (84) and formula (55) in [28]).

**5.2.1. Analysis of the boundaries of the physical acceptability of solution $\varepsilon_{KN}$ at $\rho\geq(\rho_+)_{KN}$. Calculated results**

As well as in section 3.2.1, we will consider three variants:

1. Kerr field: $\alpha_Q=0,\ \alpha_{em}=0,\ (\rho_+)_K=\alpha+\sqrt{\alpha^2-\alpha_a^2}$

$$\varepsilon_K=\frac{\alpha_a m_\varphi}{2\alpha(\rho_+)_K}. \qquad (153)$$

For the extreme Kerr field of $\left((\rho_+)_K=(\rho_-)_K=\alpha;\ \alpha^2=\alpha_a^2\right)$, solution (153) is

$$\varepsilon_K^{ext} = \frac{m_\varphi}{2\alpha}. \tag{154}$$

It follows from (153) and (154) that in presence of rotation of Kerr field source $(\alpha_a \neq 0)$, the energy of the bound fermion never achieves zero at arbitrary high values of a gravitational coupling constant $\alpha$:

$$\varepsilon_K \neq 0 \text{ at } \alpha \to \infty. \tag{155}$$

In this case, negative and zero energies of $-1 < \varepsilon_K \leq 0$ are never realized. This is possible at $m_\varphi > 0$ in (153) and (154). Thus, the bound states of fermions with $0 < \varepsilon_K < 1$ in (153) and (154) are characterized by parallel direction of the fermion spin and the angular momentum of the Kerr field source.

In order to implement the weak binding of fermion $\varepsilon_K \sim 1$, the existence of some minimal value $\alpha_{\min}$ is necessary. In the presence of event horizons, the maximal value $\alpha_a^{\max}$ is achieved for the extreme Kerr field $(\alpha_a = \alpha)$. It follows from (154) that the binding of $\varepsilon_K \sim 1$ is achieved at

$$\alpha_{\min} \sim \frac{(m_\varphi)_{\min}}{2} = 0.25. \tag{156}$$

Absence of rotation $(\alpha_a = 0)$ corresponds to the solution for the Schwarzschild field $\varepsilon_S = 0$. In this case, $\alpha_{\min}$ is also equal to $0.25$ [27].

2. The unchareged half-spin particle in the Kerr-Newman field:

$\alpha_Q \neq 0,\ \alpha_{em} = 0,\ (\rho_+)_{KN} = \alpha + \sqrt{\alpha^2 - \alpha_a^2 - \alpha_Q^2}$

$$\tilde{\varepsilon}_{KN} = \frac{\alpha_a m_\varphi}{\alpha_a^2 + (\rho_+)_{KN}^2}. \tag{157}$$

For the extreme KN field $\left((\rho_+)_{KN} = (\rho_-)_{KN} = \alpha,\ \alpha^2 = \alpha_a^2 + \alpha_Q^2\right)$, the solution of (157) is

$$\tilde{\varepsilon}_{KN}^{ext} = \frac{\alpha_a m_\varphi}{\alpha_a^2 + \alpha^2}. \tag{158}$$

In presence of rotation $(\alpha_a \neq 0)$, it follows from (157) and (158) that the energy of the bound fermion never achieves zero at an arbitrary high value of $\alpha$:

$$\tilde{\varepsilon}_{KN} \neq 0 \text{ at } \alpha \to \infty. \tag{159}$$

In this case, negative and zero energies $-1 < \tilde{\varepsilon}_{KN} \le 0$ are not realized and, hence, $m_\varphi > 0$. As in variant 1, bound states of uncharged fermions with $0 < \tilde{\varepsilon}_{KN} < 1$ are characterized by parallel direction of the fermion spin and the angular momentum of the KN field source.

It follows from (157) that inequality $\tilde{\varepsilon}_{KN} < 1$ is implemented if

$$\alpha_a m_\varphi < \alpha_a^2 + \left(\rho_+\right)_{KN}^2 . \tag{160}$$

3. The charged half-spin particle in the KN field:

$\alpha_Q \ne 0,\ \ \alpha_{em} \ne 0,\ \left(\rho_+\right)_{KN} = \alpha + \sqrt{\alpha^2 - \alpha_a^2 - \alpha_Q^2}$

$$\varepsilon_{KN} = \frac{\alpha_a m_\varphi + \alpha_{em}\left(\rho_+\right)_{KN}}{\alpha_a^2 + \left(\rho_+\right)_{KN}^2} . \tag{161}$$

For the extreme KN field $\left(\left(\rho_+\right)_{KN} = \alpha,\ \alpha^2 = \alpha_a^2 + \alpha_Q^2\right)$, the solution (161) is

$$\varepsilon_{KN}^{ext} = \frac{\alpha_a m_\varphi + \alpha_{em}\alpha}{\alpha_a^2 + \alpha^2} . \tag{162}$$

Solution (161) depends on five parameters, $\alpha_a, m_\varphi, \alpha_{em}, \alpha, \alpha_Q$, which makes its algebraic analysis difficult. The parameters $m_\varphi, \alpha_{em}$ can have different signs, and therefore, as opposed to variants 1 and 2, the entire range of bound states of $-1 < \varepsilon_{KN} < 1$ is assumed. The obvious inequality follows from here:

$$\left|\alpha_a m_\varphi + \alpha_{em}\left(\rho_+\right)_{KN}\right| < \alpha_a^2 + \left(\rho_+\right)_{KN}^2 . \tag{163}$$

As the result, we have obtained the following limitations on the physical parameters and the energy of bound fermions:

1. Kerr field: $\alpha_{em} = 0,\ \alpha_Q = 0$

$$0 < \varepsilon_K < 1,\ \ m_\varphi > 0,\ \ \alpha_a m_\varphi < 2\alpha\left(\rho_+\right)_K . \tag{164}$$

2. Uncharged particle in the Kerr-Newman field: $\alpha_{em} = 0,\ \alpha_Q \ne 0$

$$0 < \tilde{\varepsilon}_{KN} < 1,\ \ m_\varphi > 0,\ \ \alpha_a m_\varphi < \alpha_a^2 + \left(\rho_+\right)_{KN}^2 . \tag{165}$$

3. Charged particle in the Kerr-Newman field: $\alpha_{em} \ne 0,\ \alpha_Q \ne 0$

$$-1 < \varepsilon_{KN} < 1,\ \left|\alpha_a m_\varphi + \alpha_{em}\left(\rho_+\right)_{KN}\right| < \alpha_a^2 + \left(\rho_+\right)_{KN}^2 . \tag{166}$$

Table 1 presents calculated distances $\rho_m$ from the maxima of probability densities to event horizons $\left(\rho_+\right)_{KN}$ for some values of the parameters satisfying inequalities (164) - (166) with $\alpha \approx \alpha_{\min} = 0.251$. Figure 1 presents normalized probability densities (148) and integral probabilities (149) for $\alpha = 0.251$, $\alpha_Q = 0.0251$, $\alpha_a = 0.125$, $m_\varphi = 1/2$, $\alpha_{em} = 0.3;\ 0;\ -0.5$.

Table 1. Characteristics of bound fermion states with gravitational coupling constant of $\alpha = 0.251$.

| $\alpha_a$ | $\alpha_Q$ | $\alpha_{em}$ | $l, j, m_\varphi$ | $(\rho_+)_{KN}$ | $\lambda$ | $\varepsilon_{KN}$ | $\rho_m$ |
|---|---|---|---|---|---|---|---|
| 0.125 | 0 | 0 | 0, 1/2, 1/2 | 0.469 | –0.94 | 0.266 | $4.66 \cdot 10^{-2}$ |
| 0.125 | 0 | 0 | 1, 3/2, 3/2 | 0.469 | –1.9 | 0.797 | $1.12 \cdot 10^{-2}$ |
| 0.125 | 0.0251 | 0 | 0, 1/2, 1/2 | 0.467 | –0.94 | 0.267 | $4.65 \cdot 10^{-2}$ |
| 0.125 | 0.0251 | 0 | 1, 3/2, 3/2 | 0.467 | –1.99 | 0.802 | $1.12 \cdot 10^{-2}$ |
| 0.0251 | 0.125 | 0 | 0, 1/2, 1/2 | 0.467 | –0.99 | 0.057 | $4.04 \cdot 10^{-2}$ |
| 0.0251 | 0.125 | 0 | 1, 3/2, 3/2 | 0.467 | –1.99 | 0.172 | $8.54 \cdot 10^{-3}$ |
| 0.125 | 0.0251 | –0.5 | 0, 1/2, 1/2 | 0.467 | –1.02 | –0.732 | $2.78 \cdot 10^{-2}$ |
| 0.125 | 0.0251 | –0.5 | 1, 3/2, 3/2 | 0.467 | –1.995 | –0.197 | $8.55 \cdot 10^{-3}$ |
| 0.125 | 0.0251 | 0.3 | 0, 1/2, 1/2 | 0.467 | –0.89 | 0.866 | $6.91 \cdot 10^{-2}$ |
| 0.125 | 0.0251 | 0.09 | 1, 3/2, 3/2 | 0.467 | –1.88 | 0.982 | $1.18 \cdot 10^{-2}$ |
| 0.0251 | 0.125 | –0.4 | 0, 1/2, 1/2 | 0.467 | –1.01 | –0.796 | $2.58 \cdot 10^{-2}$ |
| 0.0251 | 0.125 | –0.4 | 1, 3/2, 3/2 | 0.467 | –2.01 | –0.682 | $6.68 \cdot 10^{-3}$ |
| 0.0251 | 0.125 | 0.4 | 0, 1/2, 1/2 | 0.467 | –0.98 | 0.911 | $5.76 \cdot 10^{-2}$ |
| 0.0251 | 0.125 | 0.3 | 1, 3/2, 3/2 | 0.467 | –1.98 | 0.812 | $9.71 \cdot 10^{-3}$ |
| 0.125 | 0.0251 | –0.3 | 0, 1/2, –1/2 | 0.467 | –0.97 | –0.866 | $2.24 \cdot 10^{-2}$ |
| 0.125 | 0.0251 | –0.09 | 1, 3/2, –3/2 | 0.467 | –1.93 | –0.981 | $4.26 \cdot 10^{-3}$ |
| 0.125 | 0.0251 | 0.5 | 0, 1/2, –1/2 | 0.467 | –1.1 | 0.732 | $4.02 \cdot 10^{-2}$ |
| 0.125 | 0.0251 | 0.5 | 1, 3/2, –3/2 | 0.467 | –2.05 | 0.197 | $6.73 \cdot 10^{-3}$ |
| 0.0251 | 0.125 | –0.4 | 0, 1/2, –1/2 | 0.467 | –0.99 | –0.911 | $2.23 \cdot 10^{-2}$ |
| 0.0251 | 0.125 | –0.3 | 1, 3/2, –3/2 | 0.467 | –1.99 | –0.812 | $5.85 \cdot 10^{-3}$ |
| 0.0251 | 0.125 | 0.4 | 0, 1/2, –1/2 | 0.467 | –1.02 | 0.796 | $4.85 \cdot 10^{-2}$ |
| 0.0251 | 0.125 | 0.4 | 1, 3/2, –3/2 | 0.467 | –2.02 | 0.682 | $8.79 \cdot 10^{-3}$ |

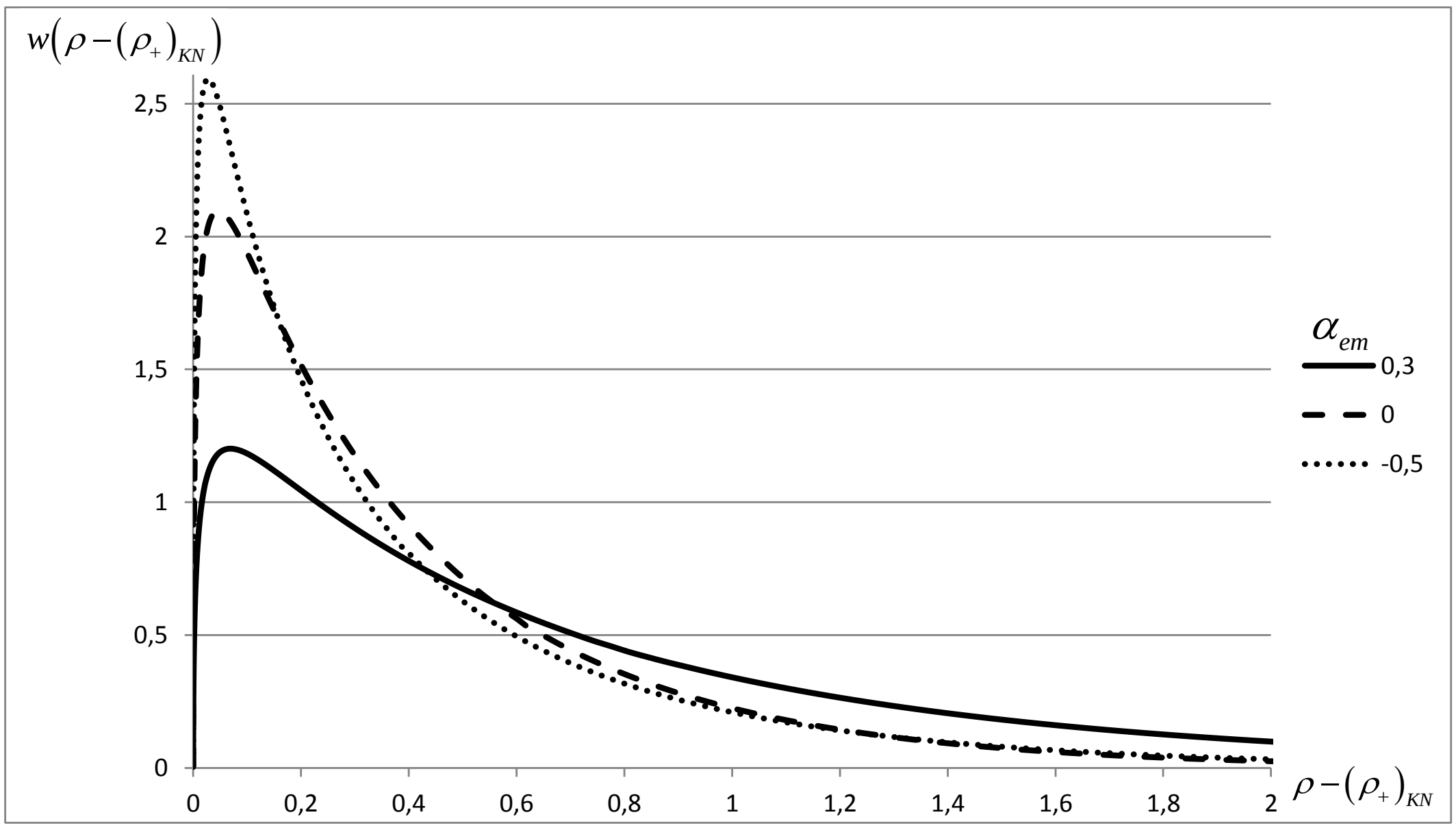

a)

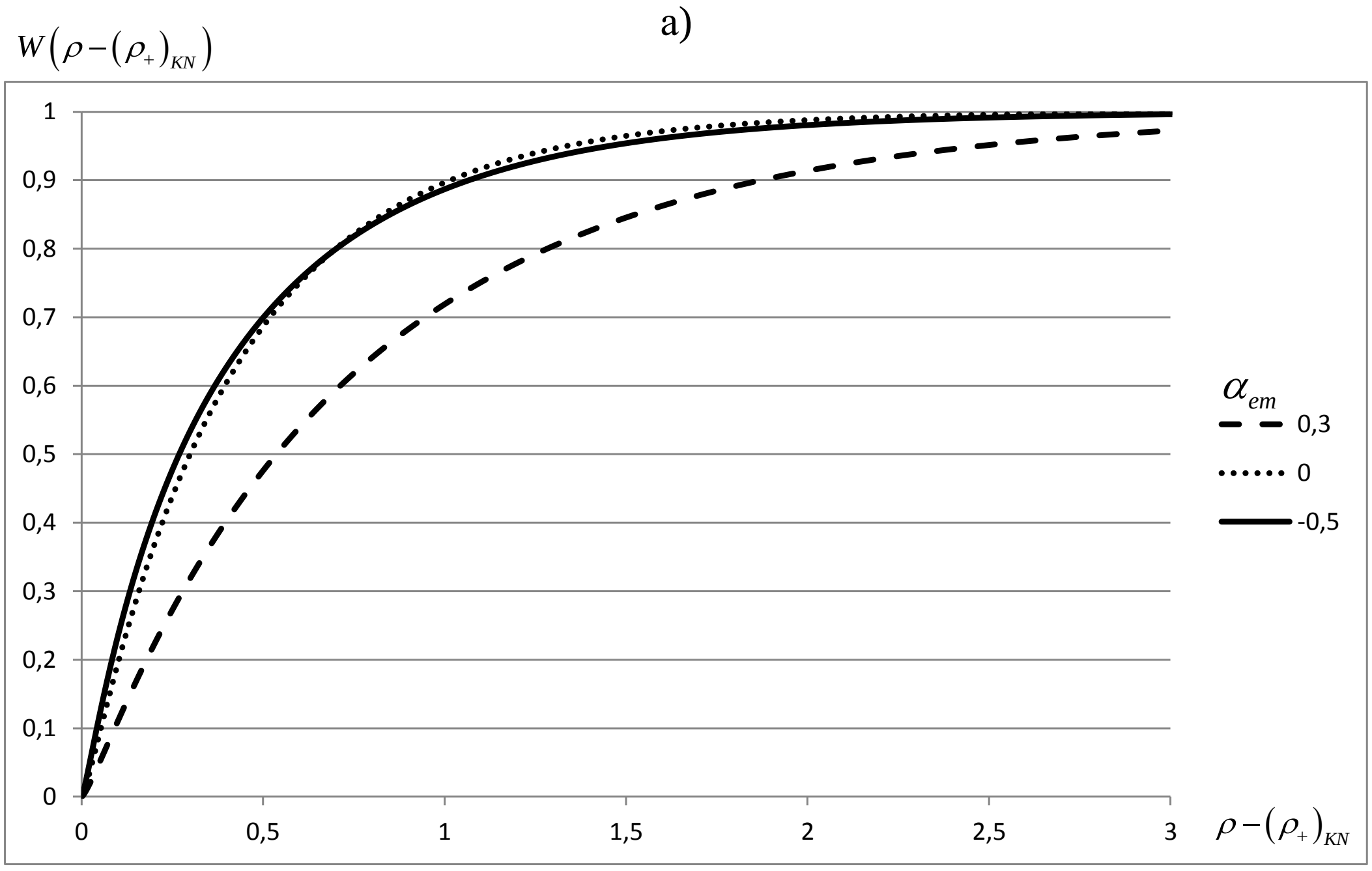

b)

Fig. 1. Curves of (a) $w\left(\rho-\left(\rho_+\right)_{KN}\right)$ and (b) $W\left(\rho-\left(\rho_+\right)_{KN}\right)$ for the bound states with $\varepsilon\left(\left(\rho_+\right)_{KN}\right)$ and $\alpha=0.251;\ \alpha_Q=0.0251;\ \alpha_a=0.125;\ m_\varphi=1/2;\ \alpha_{em}=0.3;\ 0;\ -0.5$.

**5.3. Availability of two event horizons $\left(\rho_+\right)_{KN},\left(\rho_-\right)_{KN};\alpha^2>\alpha_Q^2+\alpha_a^2$. Domain of wave functions $\rho\in\left(-\infty,\rho_-\right]$**

In this case, there exists solution (52)

$$\varepsilon_{KN}=\frac{\alpha_a m_\varphi+\alpha_{em}\left(\rho_-\right)_{KN}}{\alpha_a^2+\left(\rho_-\right)_{KN}^2}. \tag{167}$$

For the bound states of fermions $-1<\varepsilon_{KN}<1$; therefore,

$$\left|\alpha_a m_\varphi+\alpha_{em}\left(\rho_-\right)_{KN}\right|<\alpha_a^2+\left(\rho_-\right)_{KN}^2. \qquad (168)$$

Solution (167) includes the states with uncharged fermions $\left(\alpha_{em}=0\right)$ and the states with like $\left(\alpha_{em}>0\right)$ and opposite $\left(\alpha_{em}<0\right)$ charges of the KN field source and the fermion.

While determining the wave functions with known eigenvalues (167), Eq. (123) was integrated from "left to right" either from $\rho=-\infty$ with the boundary condition of (136) or from $\rho=\rho_{cl}$ with the boundary condition of (145) in the presence of the impenetrable barrier. The integration terminated at $\rho=\left(\rho_-\right)_{KN}$, in much the same way as that considered earlier for the RN field [28]. This is associated with the similar behavior of the effective potentials at $\rho\to\left(\rho_-\right)_{KN}$ and at $\rho\to\left(\rho_-\right)_{RN}$ (see (84) and (56) in [28]).

**5.3.1. Analysis of physical acceptability boundaries of solution $\varepsilon_{KN}$ at $\rho\le\left(\rho_-\right)_{KN}$. Calculated results**

Below, we will restrict ourselves to consideration of less exotic systems with the inner event horizons of $\left(\rho_-\right)_{KN}\ge 1$, i.e., with the radii greater or comparable with the Compton wavelength $l_c$ of the fermion. From the inequality of $\left(\rho_-\right)_{KN}=\alpha\left(1-\sqrt{1-\dfrac{\alpha_Q^2}{\alpha^2}-\dfrac{\alpha_a^2}{\alpha^2}}\right)\ge 1$, automatic limitations arise for value $\alpha$ at specified values of $\alpha_Q^2/\alpha^2$, $\alpha_a^2/\alpha^2$.

As above in sections 3.2.1 and 5.2.1, we will consider three variants:

1. Kerr field: $\alpha_Q=0,\ \alpha_{em}=0,\ \left(\rho_-\right)_K=\alpha-\sqrt{\alpha^2-\alpha_a^2}$

$$\varepsilon_K=\frac{\alpha_a m_\varphi}{2\alpha\left(\rho_-\right)_K}. \qquad (169)$$

It follows from (169) that $\varepsilon_K\ne 0$ at $\alpha\to\infty$, i.e., negative energies $-1<\varepsilon_K<0$ are not realized. This is possible only at $m_\varphi>0$.

2. The uncharged half-spin particle in the Kerr-Newman field:

   $\alpha_Q\ne 0,\ \alpha_{em}=0,\ \left(\rho_-\right)_{KN}=\alpha-\sqrt{\alpha^2-\alpha_a^2-\alpha_Q^2}$

$$\tilde{\varepsilon}_{KN}=\frac{\alpha_a m_\varphi}{\alpha_a^2+\left(\rho_-\right)_{KN}^2}. \qquad (170)$$

It follows from (170) that $\tilde{\varepsilon}_{KN} \neq 0$ at $\alpha \to \infty$, i.e., only the positive range of allowed energies of $0 < \varepsilon_K < 1$ is realized for bound fermions. Hence, it follows that $m_\varphi > 0$ in (170).

3. The charged half-spin particle in the KN field:

$$\alpha_Q \neq 0,\ \alpha_{em} \neq 0,\ \left(\rho_-\right)_{KN} = \alpha - \sqrt{\alpha^2 - \alpha_a^2 - \alpha_Q^2}$$

$$\varepsilon_{KN} = \frac{\alpha_a m_\varphi + \alpha_{em}\left(\rho_-\right)_{KN}}{\alpha_a^2 + \left(\rho_-\right)_{KN}^2}. \tag{171}$$

Parameters $m_\varphi, \alpha_{em}$ in (171) can have different signs and therefore the entire allowed range is acceptable for $\varepsilon_{KN}$ $\left(-1 < \varepsilon_{KN} < 1\right)$ as opposed to (169) and (170). This implies the obvious inequality

$$\left|\alpha_a m_\varphi + \alpha_{em}\left(\rho_-\right)_{KN}\right| < \alpha_a^2 + \left(\rho_-\right)_{KN}^2. \tag{172}$$

Table 2 presents the results of some demo calculations with the characteristics of bound states of fermions with the energies of (169) - (171) for the three examined variants.

Table 2. Characteristics of the bound states of fermions under the inner event horizon of the KN field with the gravitational coupling constant of $\alpha = 10$.

| | $\alpha_a$ | $\alpha_Q$ | $\alpha_{em}$ | $l, j, m_\varphi$ | $\left(\rho_-\right)_{KN}$ | $\lambda$ | $\varepsilon_{KN}$ | $\rho_m$ | $\rho_{cl}$ |
|---|---|---|---|---|---|---|---|---|---|
| particle in Kerr field | 5 | 0 | 0 | 0;1/2;1/2 | 1.34 | –0.09 | 0.093 | $8\cdot10^{-2}$ | $3.4\cdot10^{-2}$ |
| | 5 | 0 | 0 | 1;3/2;3/2 | 1.34 | –0.26 | 0.28 | $7.6\cdot10^{-2}$ | 0,104 |
| uncharged particle in Kerr-Newman field | 5 | 1 | 0 | 0;1/2;1/2 | 1.4 | –0.091 | 0.093 | $8.3\cdot10^{-2}$ | $3.6\cdot10^{-2}$ |
| | 5 | 1 | 0 | 1;3/2;3/2 | 1.4 | –0.27 | 0.278 | $7.9\cdot10^{-2}$ | 0.111 |
| | 1 | 5 | 0 | 0;1/2;1/2 | 1.4 | –0.61 | 0.169 | $8.3\cdot10^{-2}$ | 0.067 |
| | 1 | 5 | 0 | 1;3/2;3/2 | 1.4 | –1.4 | 0.508 | $7.6\cdot10^{-2}$ | 0.207 |
| charged particle in Kerr-Newman field | 5 | 1 | –1 | 0;1/2;1/2 | 1.4 | –0.335 | 0.041 | $7.1\cdot10^{-2}$ | 0.265 |
| | 5 | 1 | –1 | 1;3/2;3/2 | 1.4 | –0.515 | 0.226 | $6.7\cdot10^{-2}$ | 0.336 |
| | 5 | 1 | 0,3 | 0;1/2;1/2 | 1.4 | –0.018 | 0.108 | $8.7\cdot10^{-2}$ | –0.043 |
| | 5 | 1 | 1 | 1;3/2;3/2 | 1.4 | –0.03 | 0.33 | $9.3\cdot10^{-2}$ | –0.172 |
| | 1 | 5 | –1 | 0;1/2;1/2 | 1.4 | –0.99 | –0.304 | $8\cdot10^{-2}$ | 0.139 |
| | 1 | 5 | –1 | 1;3/2;3/2 | 1.4 | –1.8 | 0.035 | $7.2\cdot10^{-2}$ | 0.263 |
| | 1 | 5 | 1 | 0;1/2;1/2 | 1.4 | –0.25 | 0.643 | $8.7\cdot10^{-2}$ | $3.4\cdot10^{-2}$ |
| | 1 | 5 | 1 | 1;3/2;3/2 | 1.4 | –1.02 | 0.981 | $8\cdot10^{-2}$ | 0.13 |
| | 5 | 1 | –1 | 0;1/2;–1/2 | 1.4 | –3 | –0.145 | $6.5\cdot10^{-2}$ | 0.2 |
| | 5 | 1 | –1 | 1;3/2;–3/2 | 1.4 | –2.9 | –0.33 | $6.8\cdot10^{-2}$ | 0.13 |
| | 5 | 1 | 1 | 0;1/2;–1/2 | 1.4 | –3.05 | –0.04 | $7.5\cdot10^{-2}$ | $-7\cdot10^{-2}$ |
| | 5 | 1 | 1 | 1;3/2;–3/2 | 1.4 | –3 | –0.23 | $7.6\cdot10^{-2}$ | –0.376 |
| | 1 | 5 | –1 | 0;1/2;–1/2 | 1.4 | –1.12 | –0.643 | $8.6\cdot10^{-2}$ | 0.024 |
| | 1 | 5 | –1 | 1;3/2;–3/2 | 1.4 | –1.57 | –0.98 | $8.8\cdot10^{-2}$ | $8.2\cdot10^{-2}$ |
| | 1 | 5 | 1 | 0;1/2;–1/2 | 1.4 | –1.56 | 0.304 | $9.4\cdot10^{-2}$ | –0.183 |
| | 1 | 5 | 1 | 1;3/2;–3/2 | 1.4 | –2.2 | –0.035 | $9.1\cdot10^{-2}$ | –0.312 |

In Table 2, $\rho_m$ is the distance from the inner event horizon to the maximal value of probability density (148).

Figure 2 presents the normalized probability densities (148) and integral probabilities (149) for parameters $\alpha = 10, \alpha_a = 5, \alpha_Q = 1, j = 1/2, l = 0, m_\varphi = +1/2, \alpha_{em} = -1;\ 0;\ 0.3$ from Table 2.

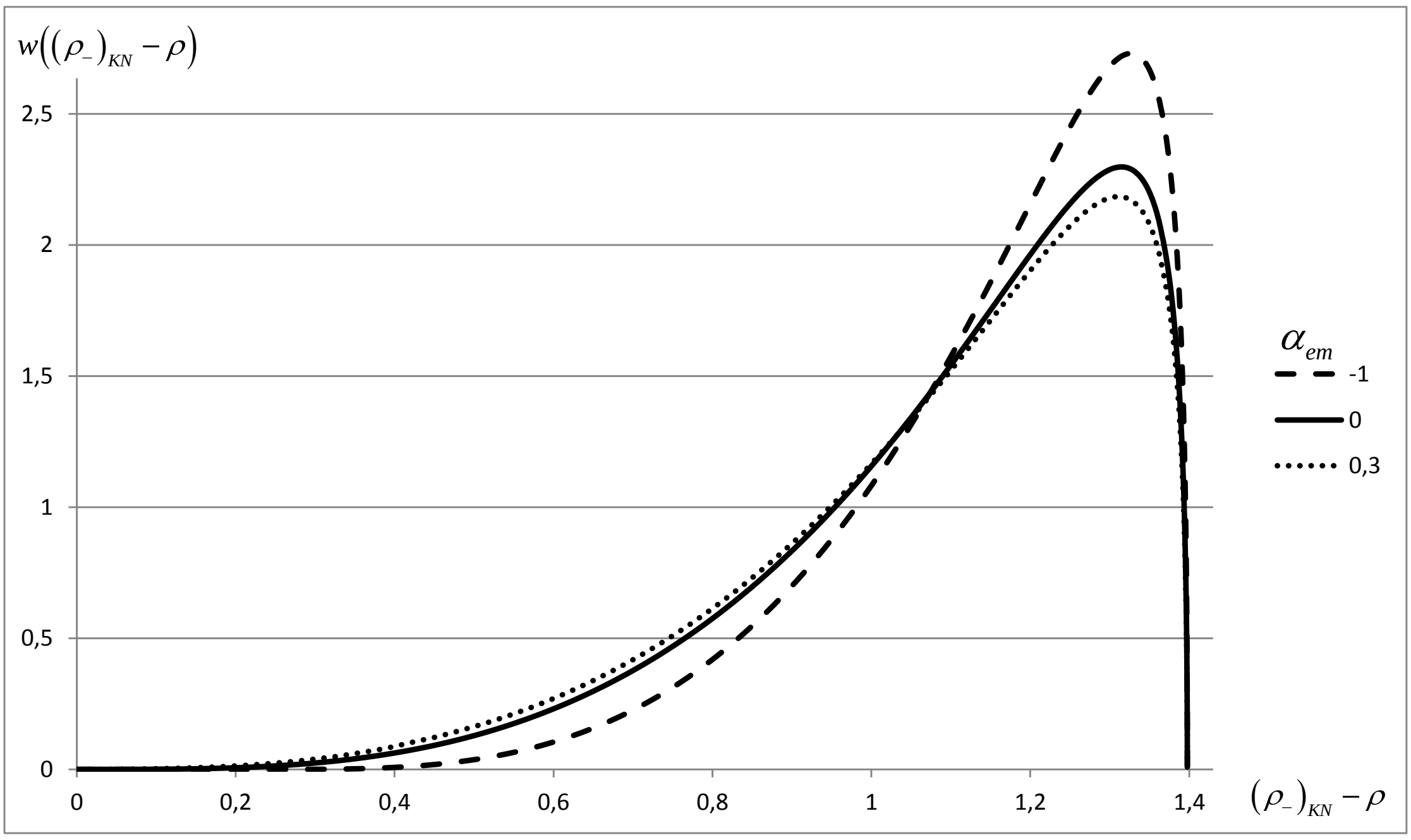

a)

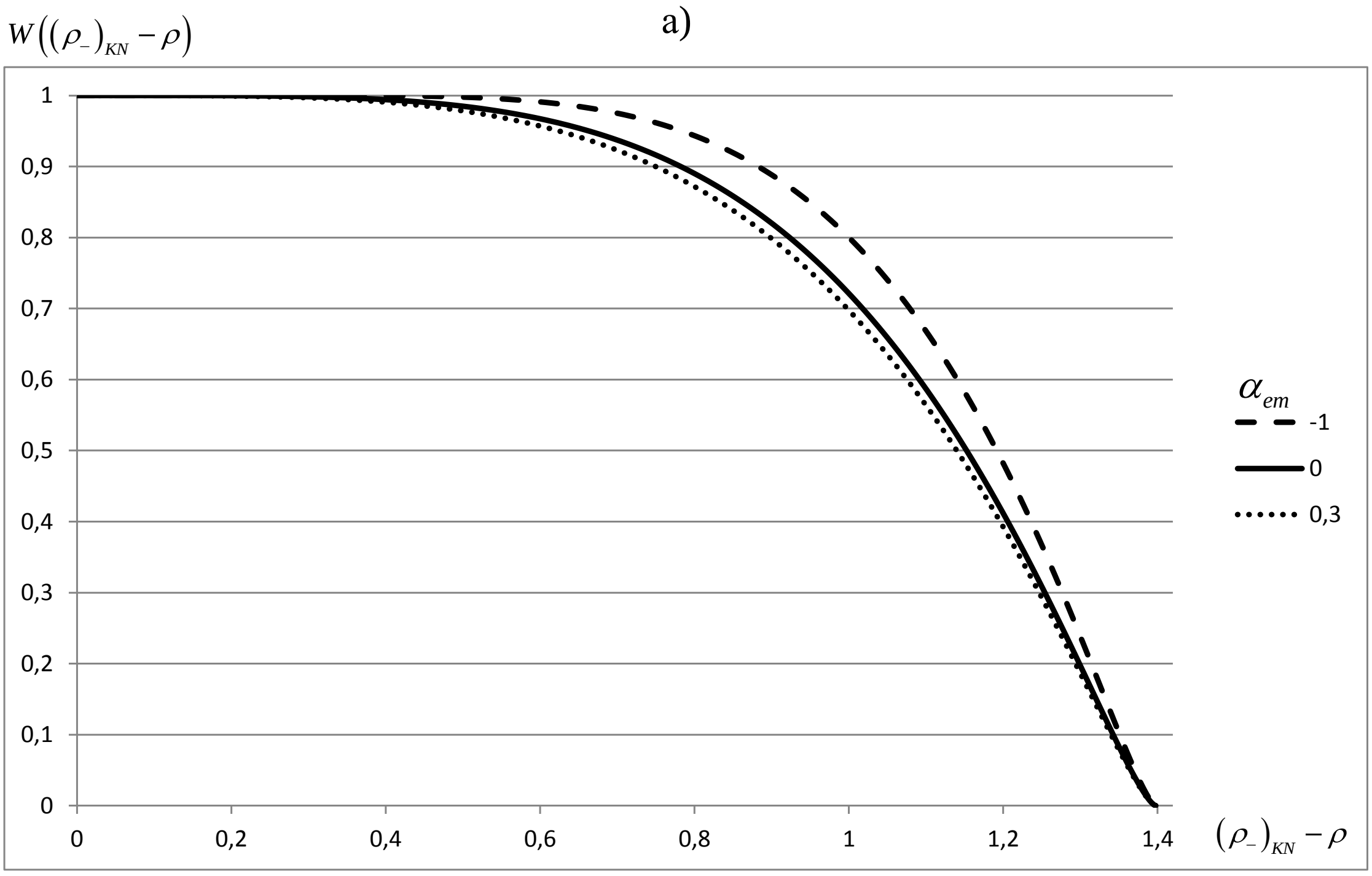

b)

Fig.2. Curves of (a) $w\left((\rho_-)_{KN} - \rho\right)$ and (b) $W\left((\rho_-)_{KN} - \rho\right)$ for bound states with $\varepsilon\left((\rho_-)_{KN}\right)$ and $\alpha = 10;\ \alpha_Q = 1;\ \alpha_a = 5;\ j = 1/2;\ l = 0;\ m_\varphi = 1/2;\ \alpha_{em} = -1;\ 0;\ 0.3$.

The calculated results demonstrate the possibility of existence of stationary bound states of fermions under inner event horizon $\rho_-$. These states with energies $\varepsilon_{KN}$ (171) are localized near the event horizon $\rho_-$. For any fermions in the Kerr field and for uncharged fermions in the Kerr-Newman field, bound states are possible only for the values of $m_\varphi > 0$ in the energy range of $0 < \varepsilon_{KN} < 1$.

There are no such limitations for charged fermions. It is seen from the calculations that the impenetrable potential barrier of type (91) with the positive or negative values of $\rho_{cl}$ exist in all the cases under consideration.

**5.4. Kerr-Newman naked singularity $\left(\alpha^2 < \alpha_a^2 + \alpha_Q^2\right)$. Domain of wave functions $\rho \in (0,\infty)$**

In contrast to sections 5.2 and 5.3, the $n, j, l$-non-degenerate discrete energy spectrum exists for fermions in the KN field of naked singularity at definite values of physical parameters.

In the numerical calculations, Eq. (123) was integrated from "left to right" (from $\rho_{\min}$ to $\rho_{\max}$). In the absence of impenetrable barrier (91), the integration begins from point $\rho_{\min} = -10^8$ with boundary condition (136). In the presence of impenetrable barrier (91), Eq. (123) is integrated from point $\rho_{\min} = \rho_{cl} + \Delta\rho_{cl}$ with boundary condition (145). A good mathematical reproducibility of the results is ensured when $\Delta\rho_{cl} = 10^{-8}$, $\rho_{\min} = -10^8$, $\rho_{\max} = 10^8$.

The energy levels in the calculations were determined at the points where function $\Phi(\varepsilon, \rho_{\max})$ varies stepwise by $\pi$ in compliance with (150).

Since effective potential (79) depends on separation constant $\lambda$, this constant is determined for each value $\varepsilon$ by solution of Eq. (128).

Let us present the results of some demo calculations to determine the discrete spectrum of $(1-\varepsilon_n)$.

**5.4.1. Kerr field $\left(\alpha_a^2 > \alpha^2,\ \alpha_Q = 0,\ \alpha_{em} = 0\right)$**

Let us consider the states of bound fermions $S_{1/2}$ $(l = 0, j = 1/2)$, $P_{1/2}$ $(l = 1, j = 1/2)$ with the parameters of $\alpha = 0.1;\ \alpha_a = 0.5;\ \alpha_a = 0.9;\ \alpha_Q = 0;\ \alpha_{em} = 0;\ m_\varphi = \pm 0.5$. The Table 3 presents the numerical calculated values of $\lambda, \rho_{cl}, 1-\varepsilon_n$.

Table 3. Numerical values of $\lambda,\ \rho_{cl},\ 1-\varepsilon_n$.

| | | $n=1,\ j=1/2,\ l=0$ | $n=2,\ j=1/2,\ l=0$ | $n=3, j=1/2,\ l=0$ | $n=2,\ j=1/2,\ l=1$ | $n=3,\ j=1/2\ l=1$ |
|---|---|---|---|---|---|---|
| $\alpha_a=0.5$ $m_\varphi=0.5$ | $\lambda$ | –0.502 | –0.5 | –0.5 | 0.915 | 0.914 |
| | $\rho_{cl}$ | 0.0023 | 0.0006 | 0.0003 | 0.0006 | 0.0003 |
| | $1-\varepsilon_n$ | $4.57\cdot10^{-3}$ | $1.2\cdot10^{-3}$ | $5.43\cdot10^{-4}$ | $1.27\cdot10^{-3}$ | $5.63\cdot10^{-4}$ |
| $\alpha_a=0.5$ $m_\varphi=-0.5$ | $\lambda$ | –1.5 | –1.5 | –1.5 | 1.23 | 1.23 |
| | $\rho_{cl}$ | –3.28 | –3.66 | –3.75 | –3.66 | –3.75 |
| | $1-\varepsilon_n$ | $5.14\cdot10^{-3}$ | $1.28\cdot10^{-3}$ | $5.66\cdot10^{-4}$ | $1.28\cdot10^{-3}$ | $5.64\cdot10^{-4}$ |
| $\alpha_a=0.9$ $m_\varphi=0.5$ | $\lambda$ | –0.1 | –0.1 | –0.1 | 0.97 | 0.97 |
| | $\rho_{cl}$ | –0.716 | –0.732 | –0.736 | –0.73 | –0.735 |
| | $1-\varepsilon_n$ | $4.36\cdot10^{-3}$ | $1.17\cdot10^{-3}$ | $5.34\cdot10^{-4}$ | $1.263\cdot10^{-3}$ | $5.61\cdot10^{-4}$ |
| $\alpha_a=0.9$ $m_\varphi=-0.5$ | $\lambda$ | –1.9 | –1.9 | –1.9 | 1.49 | 1.49 |
| | $\rho_{cl}$ | –6.54 | –7.88 | –8.27 | –7.86 | –8.27 |
| | $1-\varepsilon_n$ | $4.9\cdot10^{-3}$ | $1.25\cdot10^{-3}$ | $5.56\cdot10^{-4}$ | $1.28\cdot10^{-3}$ | $5.65\cdot10^{-4}$ |

### 5.4.2. Kerr-Newman field $\left(\alpha_a^2+\alpha_Q^2>\alpha^2\right)$

Let us consider the bound states of charged $\left(\alpha_{em}\neq0\right)$ and uncharged $\left(\alpha_{em}=0\right)$ fermions with quantum numbers $\left(l=0, j=1/2\right)$ and $\left(l=1, j=1/2\right)$ with the parameters of $\alpha=0.25$;

$$\alpha_Q=0.5;\ \alpha_a=\begin{cases}0.3\\0.7\end{cases};\ m_\varphi=\pm0.5;\ \alpha_{em}=\begin{cases}-1\\ \ 0\\0.1\end{cases}.$$

Tables 4 – 6 present the calculated results of descrete energy spectrum $1-\varepsilon_n$.

Table 4. The numerical values of $\lambda,\ \rho_{cl},\ 1-\varepsilon_n$ at $\alpha_{em}=-1$.

| | | $n=1,\ j=1/2,\ l=0$ | $n=2,\ j=1/2,\ l=0$ | $n=3, j=1/2,\ l=0$ | $n=2,\ j=1/2,\ l=1$ | $n=3,\ j=1/2\ l=1$ |
|---|---|---|---|---|---|---|
| $\alpha_a=0.3$ $m_\varphi=0.5$ | $\lambda$ | –0.785 | –0.733 | –0.717 | 0.948 | 0.939 |
| | $\rho_{cl}$ | 0.0611 | 0.047 | 0.042 | 0.044 | 0.044 |
| | $1-\varepsilon_n$ | 0.417 | 0.168 | 0.084 | 0.114 | 0.063 |
| $\alpha_a=0.3$ $m_\varphi=-0.5$ | $\lambda$ | –1.2 | –1.26 | –1.28 | 1.09 | 1.11 |
| | $\rho_{cl}$ | –0.122 | –0.144 | –0.152 | –0.148 | –0.154 |
| | $1-\varepsilon_n$ | 0.515 | 0.195 | 0.095 | 0/14 | 0.073 |
| $\alpha_a=0.7$ $m_\varphi=0.5$ | $\lambda$ | –0.45 | –0.364 | –0.334 | 0.955 | 0.944 |
| | $\rho_{cl}$ | 0.0082 | –0.04 | –0.058 | –0.051 | –0.062 |
| | $1-\varepsilon_n$ | 0.317 | 0.135 | 0.071 | 0.095 | 0.055 |
| $\alpha_a=0.7$ $m_\varphi=-0.5$ | $\lambda$ | –1.51 | –1.63 | –1.66 | 1.29 | 1.32 |
| | $\rho_{cl}$ | –0.353 | –0.446 | –0.482 | –0.463 | –0.49 |
| | $1-\varepsilon_n$ | 0.41 | 0.164 | 0.083 | 0.124 | 0.066 |

Table 5. The numerical values of $\lambda,\ \rho_{cl},\ 1-\varepsilon_n$ with $\alpha_{em}=0$.

| | | $n=1,\ j=1/2,$ $l=0$ | $n=2,\ j=1/2,$ $l=0$ | $n=3, j=1/2,$ $l=0$ | $n=2,\ j=1/2,$ $l=1$ | $n=3,\ j=1/2$ $l=1$ |
|---|---|---|---|---|---|---|
| $\alpha_a=0.3$ $m_\varphi=0.5$ | $\lambda$ | –0.704 | –0.701 | –0.7 | 0.93 | 0.929 |
| | $\rho_{cl}$ | 0.096 | 0.0943 | 0.094 | 0.094 | 0.094 |
| | $1-\varepsilon_n$ | $2.18\cdot10^{-2}$ | $6.7\cdot10^{-3}$ | $3.16\cdot10^{-3}$ | $7.75\cdot10^{-3}$ | $3.5\cdot10^{-3}$ |
| $\alpha_a=0.3$ $m_\varphi=-0.5$ | $\lambda$ | –1.295 | –1.3 | –1.3 | 1.123 | 1.124 |
| | $\rho_{cl}$ | –0.569 | –0.59 | –0.595 | –0.589 | –0.595 |
| | $1-\varepsilon_n$ | $2.38\cdot10^{-2}$ | $7.07\cdot10^{-3}$ | $3.28\cdot10^{-3}$ | $7.94\cdot10^{-3}$ | $3.55\cdot10^{-3}$ |
| $\alpha_a=0.7$ $m_\varphi=0.5$ | $\lambda$ | –0.309 | –0.303 | –0.301 | 0.93 | 0.929 |
| | $\rho_{cl}$ | –0.172 | –0.183 | –0.186 | –0.182 | –0.185 |
| | $1-\varepsilon_n$ | $2.02\cdot10^{-2}$ | $6.4\cdot10^{-3}$ | $3.07\cdot10^{-3}$ | $7.64\cdot10^{-3}$ | $3.46\cdot10^{-3}$ |
| $\alpha_a=0.7$ $m_\varphi=-0.5$ | $\lambda$ | –1.69 | –1.7 | –1.7 | 1.35 | 1.353 |
| | $\rho_{cl}$ | –0.185 | –0.206 | –0.212 | –0.204 | –0.211 |
| | $1-\varepsilon_n$ | $2.31\cdot10^{-2}$ | $6.94\cdot10^{-3}$ | $3.24\cdot10^{-3}$ | $8.06\cdot10^{-3}$ | $3.59\cdot10^{-3}$ |

Table 6. The numerical values of $\lambda,\ \rho_{cl},\ 1-\varepsilon_n$ with $\alpha_{em}=+0{,}1$.

| | | $n=1,\ j=1/2,$ $l=0$ | $n=2,\ j=1/2,$ $l=0$ | $n=3, j=1/2,$ $l=0$ | $n=2,\ j=1/2,$ $l=1$ | $n=3,\ j=1/2$ $l=1$ |
|---|---|---|---|---|---|---|
| $\alpha_a=0.3$ $m_\varphi=0.5$ | $\lambda$ | –0,.02 | –0.7 | –0.7 | 0.929 | 0.928 |
| | $\rho_{cl}$ | 0,.1 | 0.109 | 0.109 | 0.109 | 0.109 |
| | $1-\varepsilon_n$ | $8.07\cdot10^{-3}$ | $2.42\cdot10^{-3}$ | $1.13\cdot10^{-3}$ | $2.73\cdot10^{-3}$ | $1.23\cdot10^{-3}$ |
| $\alpha_a=0.3$ $m_\varphi=-0.5$ | $\lambda$ | –1.3 | –1.3 | –1.3 | 1.124 | 1.125 |
| | $\rho_{cl}$ | –0.869 | –0.896 | –0.903 | –0.895 | –0.903 |
| | $1-\varepsilon_n$ | $8.53\cdot10^{-3}$ | $2.5\cdot10^{-3}$ | $1.16\cdot10^{-3}$ | $2.75\cdot10^{-3}$ | $1.24\cdot10^{-3}$ |
| $\alpha_a=0.7$ $m_\varphi=0.5$ | $\lambda$ | –0.304 | –0.301 | –0.3 | 0.929 | 0.929 |
| | $\rho_{cl}$ | –0.216 | –0.221 | –0.223 | –0.221 | –0.222 |
| | $1-\varepsilon_n$ | $7.73\cdot10^{-3}$ | $2.36\cdot10^{-3}$ | $1.11\cdot10^{-3}$ | $2.71\cdot10^{-3}$ | $1.23\cdot10^{-3}$ |
| $\alpha_a=0.7$ $m_\varphi=-0.5$ | $\lambda$ | –1.7 | –1.7 | –1.7 | 1.353 | 1.354 |
| | $\rho_{cl}$ | –3.012 | –3.335 | –3.427 | –3.315 | –3.42 |
| | $1-\varepsilon_n$ | $8.4\cdot10^{-3}$ | $2.47\cdot10^{-3}$ | $1.15\cdot10^{-3}$ | $2.77\cdot10^{-3}$ | $1.25\cdot10^{-3}$ |

The calculated results demonstrate the existence of stationary bound states of fermions with the discrete energy spectrum in the KN field of naked singularity. As a whole, the behavior of probability densities is of the same caharcater as that when considering atomic systems in the Minkowsky space.

## 6. Cosmic censorship

The hypothesis of the cosmic censorship proposed in [54] prohibits existence in nature of singularities not covered by event horizons. However, there is still no complete proof of this hypothesis. Along with black holes, many researchers examine formation of naked singularities, their stability and distinctive features during experimental observations [55] - [60].

It is shown in [29] that there exist static metrics with time-like singularities manifesting themselves to be completely nonsingular when quantum mechanics of spinless particles are examined.

In our papers [27], [28], we validated the results of [29] as applied to motion of fermions under the horizon of the Schwarzschild metric and in the RN field of naked singularity. Indeed, the leading singularities of the effective potentials for these metrics in the neighborhood of singularity represent infinitely high potential barriers

$$U_{eff}^{S}\Big|_{\rho\to 0}=\frac{5}{8\rho^2}+\mathrm{O}\left(\frac{1}{\rho}\right),$$
$$U_{eff}^{RN}\Big|_{\rho\to 0}=\frac{3}{8\rho^2}+\mathrm{O}\left(\frac{1}{\rho}\right).$$

In these cases, the singularities at the origin in quantum mechanics are covered with repulsive barriers and the availability of singularities poses no threat to cosmic censorship.

The situation is even simplier for the KN naked singularity. The effective potential of the KN field of naked singularity is regular in the neighborhood of $\rho=0$. The point of $\rho=0$ is not a singular point in the second-order equation for fermions. The KN naked singularity in quantum-mechanic interaction with fermions poses no threat to cosmic censorship.

## 7. Conclusions

In the paper, the second-order self-conjugate equation with the effective potential is presented for the quantum- mechanical description of fermion motion in the classical Kerr-Newman field. In continuation of the previous paper [39], we note that there are no singularities in effective potential (79) and in second-order equation (76) associated with availability of ergosphere, where $g_{00}\le 0$ (in (2), the equality $g_{00}=1-\frac{r_0 r-r_Q^2}{r_K^2}=0$ determines the external and internal surfaces of the KN field ergosphere). Thus, the quantum mechanics of the Dirac equation and the second-order equation in no way highlights presence of the ergosphere with $g_{00}\le 0$.

As the result of considering the solutions of the second-order equation with the effective potentials in quantum mechanics of fermion motion in the classical Kerr-Newman field, results were obtained that qualitatively differ from the results obtained when using the Dirac equation:

1. In the presence of two event horizons $(\rho_\pm)_{KN}$, there exist regular solutions with energies

$$\varepsilon = \frac{\alpha_a m_\varphi + \alpha_{em} \left(\rho_\pm\right)_{KN}}{\alpha_a^2 + \left(\rho_\pm\right)_{KN}^2}. \tag{173}$$

These solutions represent degenerate stationary bound states of charged and uncharged fermions with square integrable wave functions and with the domains of $\rho \in \left[\left(\rho_+\right)_{KN}, \infty\right)$, $\rho \in \left(-\infty, \left(\rho_-\right)_{KN}\right]$. The wave functions weakly depend on $j, l$ and vanish on the event horizons. The fermions in bound states are localized, with the overwhelming probability, near the event horizon. The maxima of the probability densities are away from the event horizon at the distance of several fractions of the Compton wavelength of fermions.

2. In case of extreme RN fields, the fulfillment of inequality (87) actually leads to the absence of stationary bound states of fermions for any values of coupling constants.

3. For the KN field of naked singularity $\left(\alpha^2 < \alpha_Q^2 + \alpha_a^2\right)$ at definite values of physical parameters, the analysis of the effective potentials and direct numerical solutions of Schrödinger-type equation showed the existence of stationary bound states for both charged and uncharged fermions.

4. The effective potential of the KN field of naked singularity is regular in the neighborhood of $\rho = 0$. Point $\rho = 0$ is not a singular point in the second-order equation for fermions in the KN field of naked singularity. As the result, the KN naked singularity in quantum mechanical interaction with fermions does not pose any threat to cosmic censorship.

The regular solutions for the degenerate stationary bound state of fermions with energies (173) were obtained by using second-order equation (76) with effective potential (79). The wave function of Eq. (76) is related to one of the radial wave functions of the Dirac equation via non-unitary similarity transformation (72). As a result, the wave functions of the second-order equation for the degenerate stationary bound states, in contrast to the radial functions of the Dirac equation, become square integrable in the neighborhood of the event horizons of $\left(\rho_\pm\right)_{KN}$.

The self-conjugate second-order equation (76) can be obtained by squaring the covariant Dirac equation with transition from the bispinor to the spinor wave function and appropriate non-unitary similarity transformation [46]. For the plane Minkowsky time-space covarint second-order eqation for the fermions moving in external electromagnetic fields was proposed by Dirac [25].

Our examination shows (see also [27], [28]) that the use of the second-order equation expand the possibilities to obtain regular solutions of quantum mechanics of half-spin particles motion in external gravitational and electromagnetic fields.

## Acknowledgements

We express our gratitude to A.L. Novoselova for the essential technical assistance in preparation of the paper.

**APPENDIX**

## Effective potentials of gravitational and electromagnetic fields in self-conjugate second-order equations

1. Kerr-Newman field

In compliance with (68) - (71), we can obtain

$$\frac{3}{8}\frac{1}{B_{KN}^2}\left(\frac{dB_{KN}}{d\rho}\right)^2=\frac{3}{8}\left\{\frac{f_{KN}}{\omega_{KN}+\sqrt{f_{KN}}}\left[-\frac{1}{f_{KN}^2}f'_{KN}\left(\omega_{KN}+\sqrt{f_{KN}}\right)+\frac{1}{f_{KN}}\left(\omega'_{KN}+\frac{f'_{KN}}{2\sqrt{f_{KN}}}\right)\right]\right\}^2, \quad \text{(A1)}$$

$$-\frac{1}{4}\frac{1}{B_{KN}}\frac{d^2B_{KN}}{d\rho^2}=-\frac{1}{4}\frac{f_{KN}}{\omega_{KN}+\sqrt{f_{KN}}}\left[\frac{2}{f_{KN}^3}\left(f'_{KN}\right)^2\left(\omega_{KN}+\sqrt{f_{KN}}\right)-\right.$$
$$\left.-\frac{1}{f_{KN}^2}f''_{KN}\left(\omega_{KN}+\sqrt{f_{KN}}\right)-\frac{2}{f_{KN}^2}f'_{KN}\left(\omega'_{KN}+\frac{f'_{KN}}{2\sqrt{f_{KN}}}\right)+\frac{1}{f_{KN}}\left(\omega''_{KN}+\frac{f''_{KN}}{2\sqrt{f_{KN}}}-\frac{\left(f'_{KN}\right)^2}{4f_{KN}^{3/2}}\right)\right], \quad \text{(A2)}$$

$$\frac{1}{4}\frac{d}{d\rho}(A-D)=\frac{\lambda}{2}\left[\frac{1}{2}\frac{f'_{KN}}{\rho f_{KN}^{3/2}}+\frac{1}{\rho^2 f_{KN}^{1/2}}\right], \quad \text{(A3)}$$

$$-\frac{1}{4}\frac{(A-D)}{B}\frac{dB}{d\rho}=\frac{\lambda}{2\rho f_{KN}^{1/2}}\left(-\frac{f'_{KN}}{f_{KN}}+\frac{1}{\omega_{KN}+\sqrt{f_{KN}}}\left(\omega'_{KN}+\frac{f'_{KN}}{2\sqrt{f_{KN}}}\right)\right), \quad \text{(A4)}$$

$$\frac{1}{8}(A-D)^2=\frac{\lambda^2}{2f_{KN}\rho^2}, \quad \text{(A5)}$$

$$\frac{1}{2}BC=-\frac{1}{2f_{KN}^2}\left(\omega_{KN}^2-f_{KN}\right). \quad \text{(A6)}$$

In (A1) - (A6), $f_{KN}=1-\frac{2\alpha}{\rho}+\frac{\alpha_a^2+\alpha_Q^2}{\rho^2}$, $f'_{KN}\equiv\frac{df_{KN}}{d\rho}=\frac{2\alpha}{\rho^2}-\frac{2\left(\alpha_a^2+\alpha_Q^2\right)}{\rho^3}$,

$f''_{KN}\equiv\frac{d^2f_{KN}}{d\rho^2}=-\frac{4\alpha}{\rho^3}+\frac{6\left(\alpha_a^2+\alpha_Q^2\right)}{\rho^4}$, $\omega_{KN}=\varepsilon\left(1+\frac{\alpha_a^2}{\rho^2}\right)-\frac{\alpha_a m_\varphi}{\rho^2}-\frac{\alpha_{em}}{\rho}$,

$\omega'_{KN}\equiv\frac{d\omega_{KN}}{d\rho}=-\frac{2\varepsilon\alpha_a}{\rho^3}+\frac{2\alpha_a m_\varphi}{\rho^3}+\frac{\alpha_{em}}{\rho^2}$, $\omega''_{KN}\equiv\frac{d^2\omega_{KN}}{d\rho^2}=\frac{6\varepsilon\alpha_a^2}{\rho^4}-\frac{6\alpha_a m_\varphi}{\rho^4}-\frac{2\alpha_{em}}{\rho^3}$.

The sum of expressions $E_{Schr} = \frac{1}{2}\left(\varepsilon^2 - 1\right)$ and (A1) - (A6) leads to the expression for effective potential $U_{eff}^F$ (79). For the rest of the electromagnetic and gravitational fields examined in the paper, the structure of the expressions for the effective potentials does not change. Only the expressions for $f$, $f'$, $f''$, $\omega$, $\omega'$, $\omega''$ change.

2. Kerr field $\left(\alpha_Q = 0,\ \alpha_{em} = 0\right)$:

$$f_K = 1 - \frac{2\alpha}{\rho} + \frac{\alpha_a^2}{\rho^2},\ f'_K = \frac{2\alpha}{\rho^2} - \frac{2\alpha_a^2}{\rho^3},\ f''_K = -\frac{4\alpha}{\rho^3} + \frac{6\alpha_a^2}{\rho^4},\ \omega_K = \varepsilon\left(1 + \frac{\alpha_a^2}{\rho^2}\right) - \frac{\alpha_a m_\varphi}{\rho^2},$$

$$\omega'_K = -\frac{2\varepsilon\alpha_a}{\rho^3} + \frac{2\alpha_a m_\varphi}{\rho^3},\ \omega''_K = \frac{6\varepsilon\alpha_a^2}{\rho^4} - \frac{6\alpha_a m_\varphi}{\rho^4}\ .$$

3. Reissner-Nordström field $\left(\alpha_a = 0\right)$:

$$f_{RN} = 1 - \frac{2\alpha}{\rho} + \frac{\alpha_Q^2}{\rho^2},\ f'_{RN} = \frac{2\alpha}{\rho^2} - \frac{2\alpha_Q^2}{\rho^3},\ f''_{RN} = -\frac{4\alpha}{\rho^3} + \frac{6\alpha_Q^2}{\rho^4},\ \omega_{RN} = \varepsilon - \frac{\alpha_{em}}{\rho},$$

$$\omega'_{RN} = \frac{\alpha_{em}}{\rho^2}, \omega''_{RN} = -\frac{2\alpha_{em}}{\rho^3},\ \lambda = \kappa .$$

4. Schwarzschild field $\left(\alpha_Q = 0,\ \alpha_a = 0,\ \alpha_{em} = 0\right)$:

$$f_S = 1 - \frac{2\alpha}{\rho},\ f'_S = \frac{2\alpha}{\rho^2},\ f''_S = -\frac{4\alpha}{\rho^3},\ \omega_S = \varepsilon,\ \omega'_S = \omega''_S = 0,\ \lambda = \kappa .$$

5. Coulomb field (the plane space-time, $\alpha = 0,\ \alpha_Q = 0,\ \alpha_a = 0$ )

$$f_C = 1,\ f'_C = f''_C = 0,\ \omega_C = \varepsilon - \frac{\alpha_{em}}{\rho},\ \omega'_C = \frac{\alpha_{em}}{\rho^2},\ \omega''_C = -\frac{2\alpha_{em}}{\rho^3},\ \lambda = \kappa .$$